\documentclass[10pt,conference]{IEEEtran}
\IEEEoverridecommandlockouts
\usepackage{cite}
\usepackage{amsmath,amssymb,amsfonts}
\usepackage{algorithmic}
\usepackage{graphicx}
\usepackage{textcomp}
\usepackage{xcolor}
\usepackage{multirow}
\usepackage{tabularx}
\usepackage{array}
\usepackage{booktabs}
\usepackage{multirow}
\usepackage{enumitem}

\usepackage{caption}
\usepackage{subcaption}

\usepackage{xspace}

\usepackage{tablefootnote}
\definecolor{framecolor}{HTML}{E3E3E3}

\usepackage[framemethod=tikz]{mdframed}
\mdfdefinestyle{remarkstyle}{
backgroundcolor=framecolor,
roundcorner=2pt,
innerleftmargin=2pt,
innerrightmargin=2pt,
innertopmargin=2pt,
innerbottommargin=2pt,
linewidth=0.5pt,
}

\newcounter{summ}[section]

\newcommand{\mypara}[1]{\vspace{2pt}\noindent{\textit{\textbf{#1}}}\xspace}

\usepackage{wasysym}
\usepackage{amssymb}
\usepackage{pifont}

\newcounter{takeawyacount}
\def\BibTeX{{\rm B\kern-.05em{\sc i\kern-.025em b}\kern-.08em
T\kern-.1667em\lower.7ex\hbox{E}\kern-.125emX}}
\begin{document}

\title{Shedding Light on Static Partitioning Hypervisors for Arm-based
Mixed-Criticality Systems }

\author{
\IEEEauthorblockN{José Martins}
\IEEEauthorblockA{\textit{Centro ALGORITMI/LASI, Universidade do Minho} \\
jose.martins@dei.uminho.pt}
\and
\IEEEauthorblockN{Sandro Pinto} 
\IEEEauthorblockA{\textit{Centro ALGORITMI/LASI, Universidade do Minho} \\
sandro.pinto@dei.uminho.pt}}

\maketitle
\begin{abstract}
In this paper, we aim to understand the properties and guarantees of static partitioning hypervisors (SPH) for Arm-based mixed-criticality systems (MCS). To this end, we performed a comprehensive empirical evaluation of popular open-source SPH, i.e., Jailhouse, Xen (Dom0-less), Bao, and seL4 CAmkES VMM, focusing on two key requirements of modern MCS: real-time and safety. The goal of this study is twofold. Firstly, to empower industrial practitioners with hard data to reason about the different trade-offs of SPH. Secondly, we aim to raise awareness of the research and open-source communities to the still open problems in SPH by unveiling new insights regarding lingering weaknesses. All artifacts will be open-sourced to enable independent validation of results and encourage further exploration on SPH. 

\end{abstract}

\vspace{-0.125cm}

\section{Introduction} \label{sec:intro}

\par The explosion in the number of functional requirements in industries such as automotive has led to a trend for centralized architectures that consolidate heterogeneous software stacks in high-performance platforms \cite{Cerrolaza2020, Staron2021}. These typically take the form of mixed-criticality systems (MCSs) \cite{Burns2017, Esper2018} as they often integrate safety- or mission-critical workloads with real-time requirements, alongside Unix-like operating systems (OSs) providing rich functionality. Virtualization technology is the \textit{de facto} enabler for these architectures as, by definition, it allows for consolidation with strong fault encapsulation.
In this context, hypervisor design must balance, on one side, minimality for safety and security, and feature-richness and efficient sharing of resources on the other. While traditional hypervisors were optimized for the latter \cite{Hwang2008,Dall2014}, on the opposite end of the spectrum we have static partitioning hypervisors (SPH) specifically designed for MCS \cite{Ramsauer2018, Martins2020}. Besides statically assigning system resources (e.g., CPUs, memory, or devices) to virtual machines (VMs), SPH must provide latency and isolation guarantees at the microarchitectural level to comply with the freedom from interference requirements of industry safety standards such as ISO 26262 \cite{VanderLeest2015, BURGIO2017299, Cerrolaza2020, Falk2021}.

\par In this paper, we shed light on open-source SPH for Arm-based MCS. Despite the existence of multiple reports, research papers, and public artifacts, information on these systems tends to be scattered or focus on a single hypervisor or metric, while, in some cases, empirical evidence is non-existent. Thus, it is difficult to obtain a comprehensive understanding of the overall properties and guarantees of these systems in the context of MCS. To fill this gap, we conduct a leveled playing-field evaluation of four open-source static partitioning virtualization solutions, i.e., Jailhouse, Xen (Dom0-less), Bao, and the seL4 CAmkES virtual machine monitor (VMM). We drive our study based on two key requirements of modern MCS, i.e., real-time and safety, focusing on (i) performance, (ii) interrupt latency, (iii) inter-VM communication, (iv) boot time, and (v) code size. For each metric, we assess the effectiveness of the cache coloring technique, pervasive in SPH, for inter-VM interference mitigation.

\par The goal of this study is twofold.  Firstly, we aim at empowering industrial practitioners with hard data to understand the trade-offs and limits of modern Arm-based SPH as well as how best configure these systems for their use case and requirements. For example, the use of superpages significantly decreases the number of TLB misses, resulting in negligible performance overhead ($<$1\% without interference), but it is precluded by enabling page coloring, a widely adopted cache partitioning technique in these hypervisors. Also, experiments demonstrated that coloring, per se, can impact the performance up to 20\% and that it cannot fully mitigate interference, where overhead can still reach up to 60\%. Regarding inter-VM communication, we show that for bulk data transfers, buffer size choice is crucial to maximize throughput while avoiding degradation due to inter-VM interference.

\par Secondly, with the collected empirical data, we aim at raising awareness of the research and open-source communities to the still open problems in SPH, by highlighting both new and previously known weaknesses for these SPH, which seem to be mostly due to interrupt virtualization issues. Prominent examples include: (i) the need for implementing state-of-the-art mechanisms to fully mitigate inter-VM interference (e.g., memory throttling) in mainstream SPH; (ii) the extent of the impact of interference on interrupt latency, which can increase by several orders of magnitude; (iii) the lack of support for correctly handling and delivering interrupts in priority order; (iv) the absence of mechanisms that prioritize the boot of a critical VM; and (v) the lack of plasticity of the SPH architecture which might hinder achieving its own goal of allowing full IO passthrough. To address observed shortcomings, we discuss potential solutions and research directions. 

\par We made all artifacts openly available \cite{shedlight_artifact} to enable practitioners and researchers with the methods and materials to (i) independently replicate all experiments and corroborate assessed results, as well as (ii) encourage and facilitate additional experiments and further exploration of SPH.

In summary, this paper makes the following contributions: (1) presents the most comprehensive empirical study to date on popular open-source SPH focusing on a set of key metrics for modern MCS; (2) provides hard empirical data to empower industrial practitioners with the knowledge to understand the limits and trade-offs of SPH; (3) raises awareness of the research and open-source communities to the open problems of SPH by shedding light on their shortcomings; and finally, (4) opens all artifacts to enable independent validation of results and facilitate further research.


\section{Background}
\label{sec:back}

In this section, we start by overviewing  Armv8-A virtualization support. We then explain the concept of static partitioning virtualization, including key techniques implemented in SPH. Finally, we describe Xen, Jailhouse, Bao, and seL4 CAmkES.

\subsection{Arm Virtualization}
\label{sec:arm-virt}


\mypara{CPU \& Memory.} Given the widespread proliferation of virtualization in the last decades, Arm implemented hardware support since version 7 of the ISA. The most recent version of the architecture, i.e., Armv8/9-A, extends the privileged architecture with a dedicated hypervisor privilege mode (EL2) which sits between the secure firmware mode (EL3) and the kernel/user modes (EL1/EL0) \cite{arm_virt} where guests execute. A hypervisor running at EL2 has fine-grained control over which CPU resources are directly accessible by guests (e.g., control registers). Access to a denied functionality by a guest OS results in a trap to the hypervisor. It is possible to route specific guest exceptions and system interrupts to EL2. Other resources that can be managed by the hypervisor include the CPU-private generic timer and the performance monitor unit (PMU). 
EL1/EL0 memory accesses are subject to a second stage of translation which is in full control of the hypervisor \cite{arm_virt}. Any guest access to a memory region not mapped in the second stage of translation will result in a precise trap to EL2. Arm provides multiple ``translation granules", resulting in pages of different sizes: 4 KiB, 16 KiB, and 64 KiB. For each page size it is also possible to map large contiguous memory regions. These are known as superpages (or hugepages), which reduces TLB pressure. The more commonly used 4KiB granule allows for 1GiB and 2MiB superpages. Arm also defines the System Memory-Management Unit (SMMU), that extends memory virtualization mechanisms from the CPU to the bus, to restrict VM-originated direct-memory accesses (DMAs). 

\mypara{Interrupts.} Arm virtualization acceleration spans the full platform, including the Generic Interrupt Controller (GIC). The GICv2 standard has two main components: a central distributor and a per-core interface. All interrupts are routed first to the distributor, which then forwards them to the interfaces. The distributor allows the configuration of interrupt parameters (e.g., priority, target CPU) and the monitoring of interrupt state, while the interface enables the core management of interrupts. GICv2 provides virtualization support only on the interface; there is a fully virtual interface with which the guests can directly interact without VM exits. The distributor, however, must be fully emulated. Furthermore, all interrupts must first be handled by the hypervisor, which can then inject them in the VM, by writing to GIC list registers (LRs). These registers essentially take the place of the distributor for the virtual interface: when a given interrupt (along with metadata such as priority or state) is present on a register, it is forwarded to the virtual interface. The GICv2 spec limits the number of LRs to a maximum of 16. GICv3 and GICv4 provide support for direct delivery of hardware interrupts to VMs; however, this feature is only implemented for inter-processor interrupts (IPIs) and message-signaled interrupts (MSIs), i.e., interrupts implemented as write operations to special interrupt controller registers and propagated via the system interconnect. Standard wired interrupts, propagated by dedicated side-band signals, are still subject to the mentioned limitation, i.e., hypervisor interrupt injection through the list register.

\subsection{Static Partitioning Virtualization (SPV)}

Static partitioning is the practice of, either at build or initialization time, distributing all platform resources to different subsystems. This can be materialized in many shapes and forms, depending on the hardware primitives. Virtualization is a natural enabler for the static partitioning architecture, due to the strong encapsulation guarantees and flexible resource assignment.
Hypervisors designed for the static partitioning use case (or providing such a configuration) have three fundamental properties: (i) exclusive assignment of virtual CPUs to physical CPUs (i.e., no scheduler); (ii) static allocation, assignment, and mapping of all hypervisor and VM memory at build or initialization time; and (iii) direct assignment of devices to VMs (passthrough) and exclusive allocation of their interrupts to the same VM. To implement this efficiently, these hypervisors are highly dependent on virtualization hardware support both at the CPU and platform level (e.g., SMMU). SPH also have non-functional requirements centered around minimizing interrupt latency and inter-VM interference. Thus, over the past few years, there have been efforts to enhance SPH with mechanisms to address these requirements. These include cache coloring and, analogously to what has been done for x86 \cite{able2012}, direct injection in Arm processors.
Furthermore, it is important for the code base to be minimal and follow industry coding standards (e.g., MISRA); this eases functional safety (FuSa) certification efforts. 

\mypara{Cache Coloring.} In SPH, VMs still share microarchitectural resources such as the last-level cache (LLC). The behavior and memory access pattern of one VM might result in the eviction of another VM's cache lines, impacting the latter's hit rate and consequently its execution time. Thus, there is the need to partition shared caches, assigning each partition to a different VM. While in the past, Armv7 processors provided hardware means to apply this partitioning by way of per-master cache-locking, modern-day Arm CPUs do not provide those facilities. A solution is cache coloring, a software technique for index-based cache partitioning \cite{Gracioli2015}. Cache coloring explores the intersection of the virtual addresses' cache index and page number when creating virtual-to-physical memory mappings. Each color is a specific bit pattern in this intersection that maps only to specific cache sets. Thus, hypervisors can control which cache sets are assigned to a given VM by selecting which physical pages are mapped to it. By exclusively assigning a cache partition (i.e., group of cache sets or colors) to a given VM, cache coloring fully eliminates the conflict misses resulting from inter-VM contention. Cache coloring can also be applied to the hypervisor itself by assigning it one or more specific colors.



\mypara{Direct Interrupt Injection.} Direct interrupt injection is a new technique implemented in Arm-based SPH to eliminate the need of the hypervisor mediating interrupt injection. With this technique, the hypervisor passes through the physical GIC CPU interface and routes all interrupts directly to the VM by configuring the CPU to trigger interrupt traps directly at EL1, i.e., kernel mode. The hypervisor must still emulate the shared distributor to ensure isolation between VMs, i.e., prevent misconfiguration of a given VM's interrupts by another VM. This allows physical interrupts to be directly delivered to the VM with no hypervisor intervention, reducing latency to native execution levels. The forfeiting of interrupts should not be a major issue as SPH do not directly manage devices. However, SPH still need to communicate internally using IPIs. Direct interrupt injection implementations address this issue by leveraging standard software-delegated exception interface (SDEI) \cite{arm_sdei} events instead of directly using IPIs. SDEI is implemented by firmware, allowing the hypervisor to register an event during initialization. The hypervisor can then trigger the event by issuing a system call to firmware (via a secure monitor call instruction, SMC), which will result in diverting execution to a predefined hypervisor handler, similarly to Unix signals. 
In reality, firmware maps these events to its own secure reserved IPIs since, as part of TrustZone [16], the GIC provides further facilities to reserve interrupts to EL3.


 \begin{figure*}[!t]
    \centering
    \includegraphics[width=0.9\textwidth]{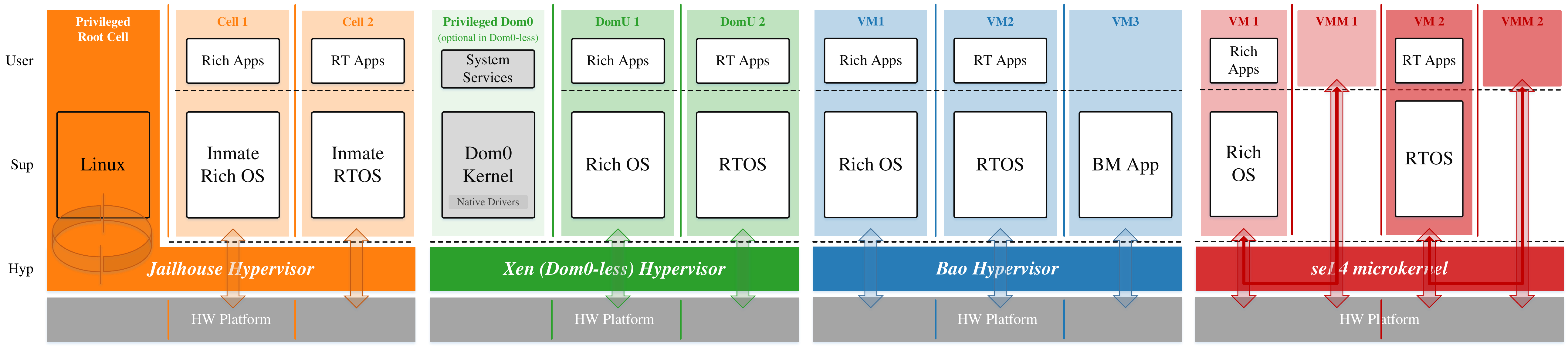}
    \caption{Architectural overview of the assessed hypervisors: Jailhouse, Xen (Dom0-less), Bao and seL4 CAmkES VMM}
    \label{fig:hypervisors}
    \vspace{-0.50cm}
\end{figure*}

\subsection{Static Partitioning Hypervisors (SPH)}


\mypara{Jailhouse Hypervisor.}  Jailhouse \cite{Ramsauer2017, Ramsauer2018} is an open-source hypervisor developed by Siemens. Unlike traditional baremetal hypervisors, Jailhouse leverages the Linux kernel to boot and initialize the system and uses a kernel module to install the hypervisor. Once Jailhouse is activated, it runs as a baremetal component, taking full control over the hardware. Jailhouse has no scheduler and only leverages the ISA virtualization primitives to partition hardware resources across multiple isolated domains, a.k.a. ``cells". Guest OSes or baremetal applications running inside cells are called ``inmates".
The mainline includes support for x86 and Armv7/8-A, and a work-in-progress RISC-V port\cite{Ramsauer2022}. 
The research community has been actively contributing with mechanisms to enhance predictability, namely: cache coloring, DRAM bank partitioning \cite{Kloda2019}, memory throttling, and device quality of service (QoS) regulation \cite{Sohal2020}. An unofficial fork including these features is available \cite{jailhousert}. Direct injection \cite{Biondi2021} was also implemented. 

\mypara{Xen (Dom0-less) Hypervisor.} Xen \cite{Hwang2008} is an open-source hypervisor widely used in a broad range of application domains. A key distinct feature of Xen is its dependency on a privileged VM (Dom0) that typically runs Linux, to manage non-privileged VMs (DomUs) and interface with peripherals. Xen was initially designed for servers and desktops, but has found also adoption on embedded applications. For embedded and automotive applications, Xilinx has led the implementation of Xen Dom0-less. With this novel approach, it is possible to have a Xen deployment without any Dom0, booting all guests directly from the hypervisor and statically partitioning the system.
A patch for guest and hypervisor cache coloring support \cite{Corradi2020} is available. There is also a SIG working towards facilitating downstream FuSa certifications by fostering multiple initiatives within the community including MISRA refactoring, or providing the option of running Zephyr \cite{zephyr_project_2023} as Dom0. Besides Armv8-A, Xen also supports x86, and Armv8-R and RISC-V ports are underway.

\mypara{Bao Hypervisor.} Bao \cite{Martins2020} is an open-source static partitioning hypervisor that was made publicly available in 2020. It implements the pure static partitioning architecture, i.e., a minimal, thin-layer of privileged software which leverages the existing ISA virtualization primitives to partition the hardware. Bao has no scheduler and does not rely on any external libraries or privileged VM (e.g., Linux), consisting on a standalone component which depends only on standard firmware to initialize the system and perform platform-specific tasks such as power management. 
Bao originally targeted Armv8-A \cite{Martins2020}. The mainline now includes support for RISC-V \cite{Sa2021}, Armv7-A, and Armv8-R ports are in the making. 
Bao was specifically designed to provide strong real-time and safety guarantees. It implements hardware partitioning mechanisms to guarantee true freedom from interference, i.e., cache coloring (VM and hypervisor), and direct interrupt injection. There are ongoing efforts to implement memory throttling. 

\mypara{seL4 CAmkES VMM.} seL4 is a formally verified microkernel \cite{Klein2009}. 
Its design model revolves around the use of capabilities.
When used as a hypervisor, seL4 executes in hypervisor mode (e.g, EL2) and exposes extra capabilities and APIs to manage virtualization functionality \cite{Heiser2020_1}. A user-level VMM uses its resource capabilities to create VMs. 
As of this writing, only the seL4 CAmkES VMM \cite{Klein2018, millwood2020} code is open-source.
Each CAmkES VMM manages a single VM. One current issue of the CAmkES VMM is that, although it supports multicore VMs, each VMM runs as a single thread pinned to a single CPU.
seL4 supports x86, Armv7/8-A and RISC-V, but the latter is not supported by CAmkES VMM.
In CAmkES, resources are statically allocated to each component using capabilities. 
Originally, seL4 provided only a priority-based preemptive scheduler.
The newest MCS kernel extends it with scheduling context capabilities, allowing time management policies to be defined in user space \cite{Lyons2018}. 
Cache coloring has also been implemented in seL4 \cite{Ge2019}, not only at the user/VM level, but also for the kernel, but it was not publicly available at the time of writing.
seL4 has formal proofs for its specification, implementation from C to binary, and security properties \cite{Murray2013, Klein2014}. 
There are also ongoing efforts to extend the formal verification to prove the absence of covert timing channels \cite{Heiser2020}. 
Finally, CAmkES is being deprecated in the near future in favor of the seL4 Core Platform (seL4CP) \cite{Heiser2022}. seL4CP will also provide support for per-VM user-mode VMMs\footnote{Only after the bulk of this work was carried out, virtualization support in seL4CP was made openly available. At the time of writing, it still appears to be in a beta stage and not as mature as CAmkES.} while promising to alleviate the performance overhead of CAmkES.

\section{Methodology and Experimental Setup} \label{sec:meth-setup}


\subsection{Methodology} \label{subsec:meth}
\vspace{-0.13cm}
\mypara{Selected Hypervisors.} We have selected four open-source SPH (Fig.\ref{fig:hypervisors}). Jailhouse and Bao were designed for the static partitioning use case; both are open-source and target Arm platforms. Xen Dom0-less is a novel deployment that allows directly booting multiple VMs (bypassing Dom0) and passthrough of peripherals to VMs. Finally, seL4 is a well-established open-source microkernel, which can be used as a hypervisor in combination with a user-level VMM. The seL4 CAmkES VMM is an open-source reference VMM implementation with static allocation of resources. These systems are actively maintained, adopted for commercial purposes, and there is a fair amount of information about them. We have excluded other open-source SPH that do not support Armv8-A such as the SPH architecture pioneer Quest-V \cite{West2014, West2016, sinha2021}, and ACRN \cite{li2019}, as well as open-source hypervisors that don't explicitly target static partitioning (e.g., KVM \cite{Dall2014}, Xvisor \cite{Patel2015}). We have excluded microkernels such as NOVA \cite{Steinberg2010} due to the lack of availability of an open-source reference user-space VMM, and because we believe seL4 serves as a faithful representative of the microkernel architecture. TrustZone-assisted hypervisors \cite{Pinto2017, Martins2017, Pinto2019a} were left out due to multicore scalability issues and lack of active maintenance.  
Finally, we have excluded commercial products (e.g., PikeOS, LynxSecure) as these often require licenses the authors did not have access to, and that would limit wide access to the study artifacts.

\mypara{Empirical Evaluation.} The evaluation focuses on performance, interrupt latency, inter-VM communication latency and bandwidth, boot time, and code size. We also assess the effect of interference and of the available mitigation mechanism (i.e., cache coloring). Although we consider virtual device performance, IO interference, and applied security techniques such as stack canaries or guards, data execution prevention or control-flow integrity very relevant, these are out of scope of this work. We advocate for a follow up study as future work.

\subsection{Experimental setup} \label{subsec:setup} 
\vspace{-0.15cm}
\mypara{Hardware Platform.} Experiments were carried out on a Xilinx ZCU104, featuring a Zynq Ultrascale+ SoC. It includes a quad-core Cortex-A53 running at 1.2 GHz, a GIC-400 (GICv2) featuring 4 list registers, and an  MMU-500 (SMMUv2).
Cores have private 32KiB separate L1 instruction and data caches, and share a L2 1MiB unified cache. It also includes a programmable logic (PL) component (i.e., FPGA).

\mypara{Hypervisors configuration.} We made an effort to use the latest versions of each SPH. Still, we applied a few patches to Jailhouse, Xen, and Bao to include features such as coloring or direct injection, which are not yet fully merged. Further, we had to make small adjustments to all SPH to enable homogeneous configurations (e.g., uniforming VM memory map), allow direct guest access to PMUs, or instrumenting hypervisors for specific experiments. For each SPH, we leveraged the default configuration for the target SoC, with some tweaked options such as disabling debug and logging features. There were, however, specific adjustments that were made on a per-hypervisor basis. For example, to remove or minimize the invocation of a scheduler in Xen, we used the null scheduler and disabled trapping of wait-for-interrupt (WFI) instructions; in seL4, since it was not possible to disable the timer tick, we configured the tick with a period of about 5 seconds. We compiled all hypervisors with GCC 11.2, with the default optimization level defined by each hypervisor's build system. All these SPH configurations and modifications are available and clearly discernible in the provided artifact \cite{shedlight_artifact}.

\mypara{VM configuration.} VM configurations are as similar as possible, mainly w.r.t. number of vCPUs and memory. For Jailhouse and seL4-VMM, where memory must be manually allocated, we set memory regions aligned to 2 MiB. The only device assigned to each VM is a UART. We evaluated two different classes of VMs: (i) large VMs running Linux (v5.14), as representative of rich, Unix-like OSs; and (ii) small VMs running baremetal applications or FreeRTOS (v10.4), as representative of critical workloads. When cache coloring is enabled, we assign half of the colors (four out of eight\footnote{We consider only eight cache colors while, in truth, the target platform allows for 16. We do this to avoid color assignment configurations that would partition the L1 cache.}) to the VM executing the benchmark, three colors to the interference application, and one color to the hypervisor (just supported in Bao and Xen). Note that color assignment configuration can significantly impact the final measurements for all metrics. In real deployments, the color assignment should be carefully defined based on the profile of the final system.

\mypara{Interference Workload.} When evaluating memory hierarchy interference, we use a custom baremetal guest which continuously writes a buffer with the size of the LLC (1MiB). Unless noted otherwise, this interference guest runs on a VM with two vCPUs. We stress that although parameterized to cause a significant level of interference, the observed effects caused by the interference workload do not necessarily reflect the worst case that could be achieved if further fine-tuned.

\mypara{Measurement tools.} We use the Arm PMU to collect microarchitectural events on benchmark execution. The selected events include instruction count, TLB accesses and refills, cache access and refills, number of exceptions taken, and number of interrupts triggered; we register the exception level on which these events occur. For the Linux VMs, we use the perf tool \cite{perf} to measure the time and to collect microarchitectural events. For baremetal or RTOS VMs, we use the Arm Generic Timer, with a resolution of 10 ns, and a custom PMU driver.


\subsection{Threats to validity} \label{subsec:threads} 

Experiments were independently conducted by two researchers. Each used a different ZCU104 platform and pre-agreed VM configurations (cross-checked). We have contacted key individuals and/or maintainers as representatives of each SPH community. We have received replies from all of them, which led to a few iterations and repetition of some experiments. Overall, the comments and issues raised by these individuals are reflected in the presented ideas and results. Despite all efforts, these experiments may still be subject to latent inaccuracies. We will open source all artifacts to enable independent validation of the results.
This study may also include limitations on the generalization to other platforms. For the hardware platform, we argue both the SoC (Zynq Ultrascale+) and the Cortex-A53 are representative of others used in automotive and industrial settings (e.g., NXP i.MX8 or Renesas R-Car M3). To corroborate this, we have also carried out the performance and interrupt latency experiments for the Bao hypervisor in an NXP i.MX8QM, which features the GIC-500 (GICv3). The obtained results are fully consistent with those presented in Sections \ref{sec:perf} and \ref{sec:irq-latency}. Furthermore, we argue next generation platforms, such as i.MX9 featuring Cortex-A55 CPUs, implement very similar microarchitectures.
\begin{figure*}[t]
    \centering
    \includegraphics[width=0.95\textwidth]{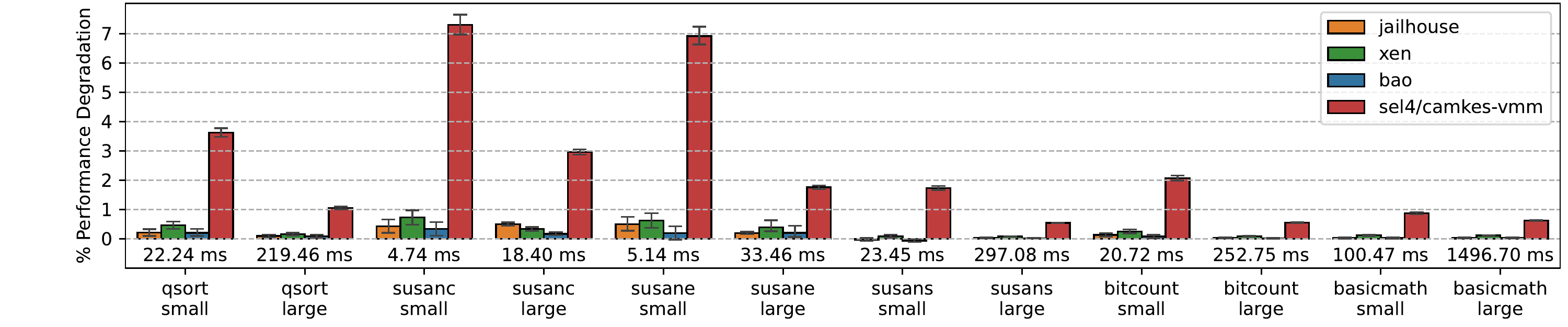}
    \vspace{-0.2cm}
    \caption{Relative performance degradation for the MiBench Automotive and Industrial Control Suite.}
    \label{fig:mibench}
    \vspace{-0.5cm}
\end{figure*}


\section{SPH: Performance} \label{sec:perf}

\begin{figure}[!t]
    \centering
    \begin{subfigure}{0.5\textwidth}
        \centering
        \begin{subfigure}{0.49\textwidth}
            \centering
            \includegraphics[width=1\textwidth]{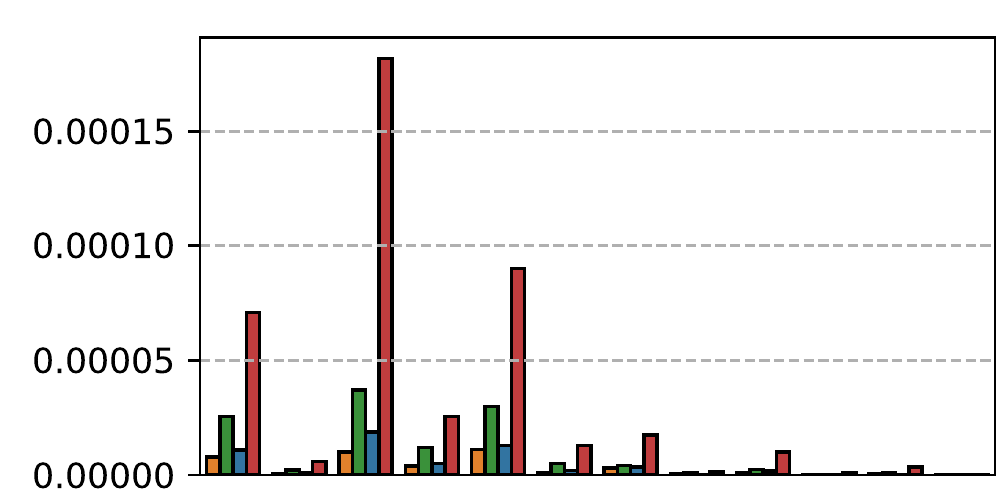}
            \vspace{-0.6cm}
            \caption{Hyp. L2 cache miss per instr.}
            \label{fig:mibench-hyp-l2-miss}
        \end{subfigure}
        \begin{subfigure}{0.49\textwidth}
            \centering
            \includegraphics[width=1\textwidth]{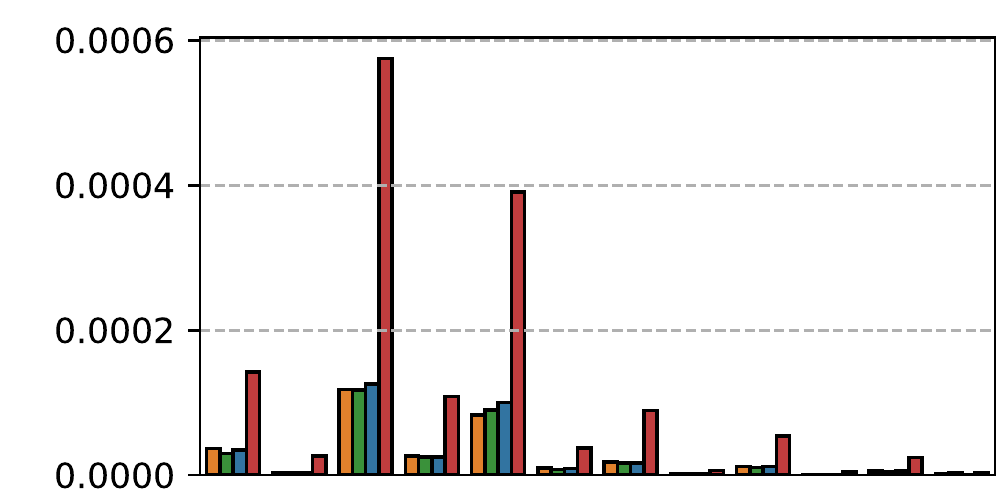}
            \vspace{-0.6cm}
            \caption{Guest iTLB miss per instr.}
            \label{fig:mibench-guest-itlb-miss}
        \end{subfigure}
    \end{subfigure}
    \begin{subfigure}{0.5\textwidth}
        \centering
        \begin{subfigure}{0.49\textwidth}
            \centering
            \includegraphics[width=1\textwidth]{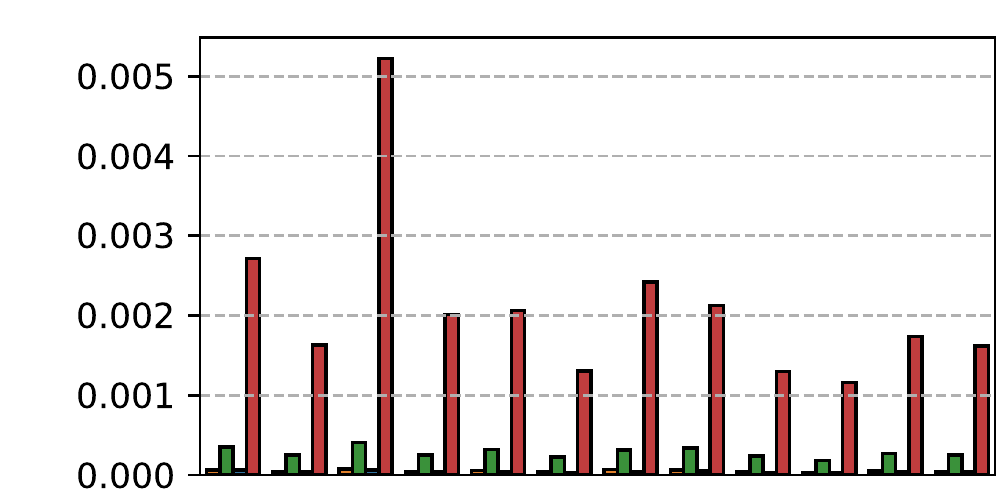}
            \vspace{-0.6cm}
            \caption{Hyp./Guest instr. ratio}
            \label{fig:mibench-inst-ratio}
    \end{subfigure}
        \begin{subfigure}{0.49\textwidth}
            \centering
            \includegraphics[width=1\textwidth]{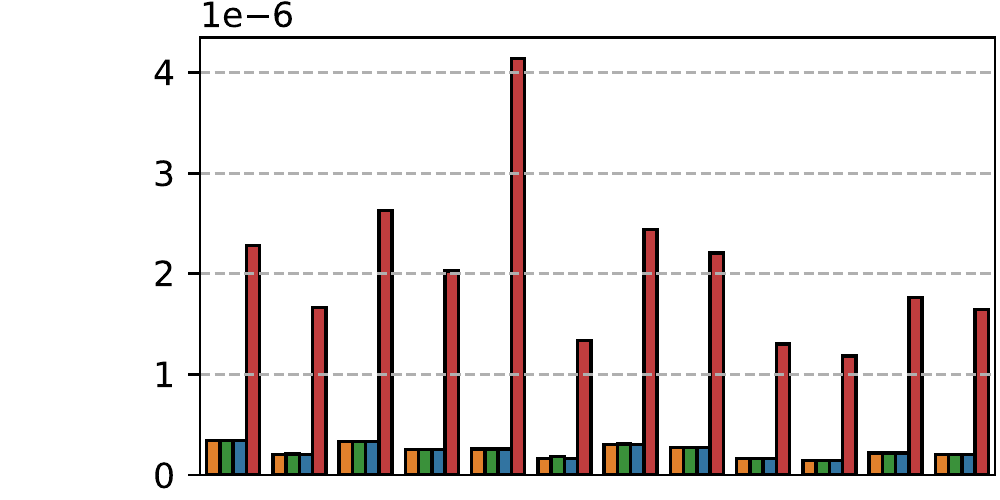}
            \vspace{-0.6cm}
            \caption{Hyp. exceptions per instr.}
            \label{fig:mibench-hyp-excp}
        \end{subfigure}
    \end{subfigure}
    \caption{MiBench AICS microarchitectural events.}
    \label{fig:mibench-events}
    \vspace{-0.25cm}
\end{figure}

\begin{figure}
    \centering
    \begin{subfigure}{0.5\textwidth}
        \centering
        \begin{subfigure}{0.49\textwidth}
            \centering
            \includegraphics[width=1\textwidth]{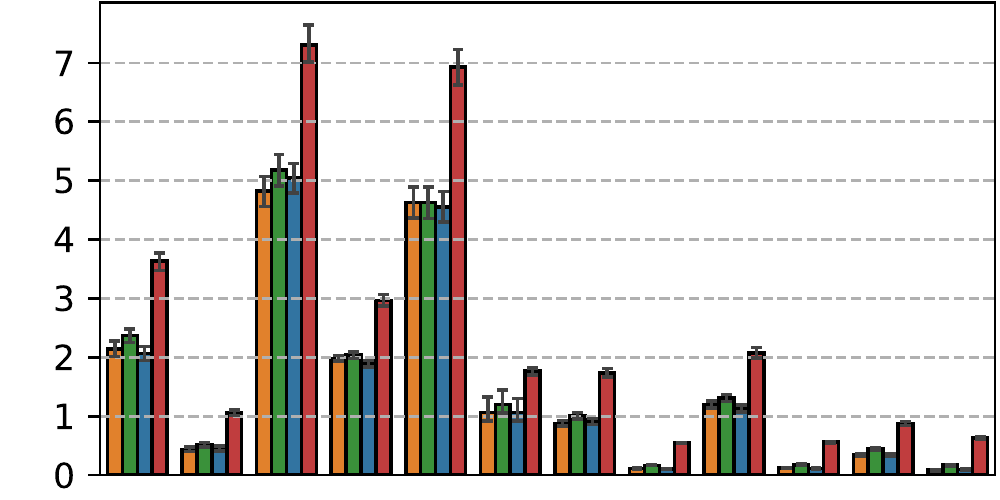}
            \vspace{-0.6cm}
            \caption{\% Performance Degradation}
            \label{fig:mibench-nosuper-inst-ratio}
    \end{subfigure}
        \begin{subfigure}{0.49\textwidth}
            \centering
            \includegraphics[width=1\textwidth]{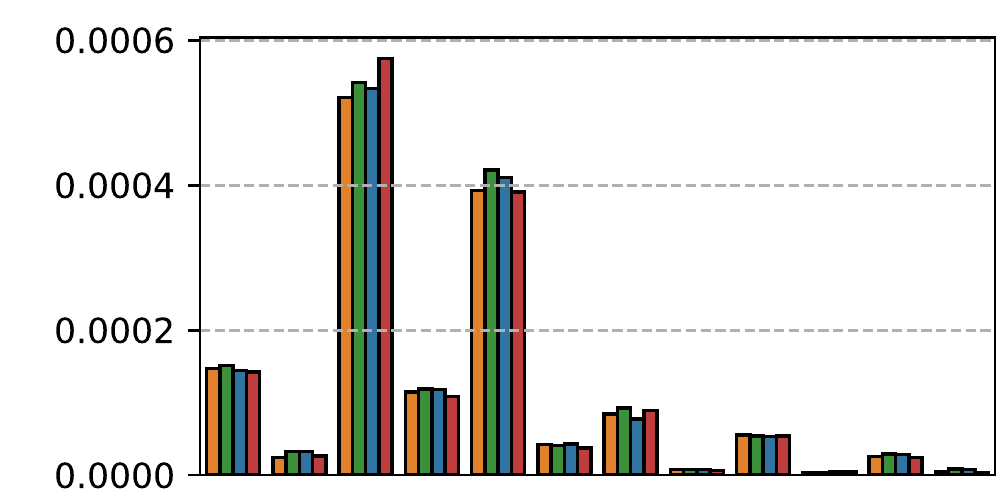}
            \vspace{-0.6cm}
            \caption{Guest iTLB miss per instr.}
            \label{fig:mibench-nosuper-hyp-excp}
        \end{subfigure}
    \vspace{0cm}
    \end{subfigure}
    \caption{MiBench AICS without the use 
    of superpages on second-stage translation.}
    \label{fig:mibench-nosuper}
    \vspace{-0.7cm}
\end{figure}

\par We start by assessing the performance degradation\footnote{Performance degradation is the ratio between the total execution time of the benchmark running atop the hypervisors and native execution.} of a single-core Linux VM atop each SPH. The main results are depicted in Figures \ref{fig:mibench}, \ref{fig:mibench-events}, and \ref{fig:mibench-nosuper}.  We then evaluate the system under interference to understand the effectiveness of microarchitectural isolation mechanisms available in each SPH.

\mypara{Selected Benchmark.} We use the MiBench Embedded Benchmarks' Automotive and Industrial Control Suite (AICS)\cite{mibench}. These benchmarks are intended to emulate the environment of embedded applications such as airbag controllers and sensor systems. Each test has two variants: \textit{small} operates in a reduced input data set representing a lightweight use of the benchmark, while \textit{large} operates over a considerable  input data set, emulating a real-world application scenario.

\mypara{Base Performance Overhead.} Fig. \ref{fig:mibench} presents the relative performance degradation for the MiBench AICS. For each benchmark, below the plotted bars, we present the average absolute execution time for the native execution. The first observation is that, independently of the hypervisor, different benchmarks are affected to different degrees. Secondly, Jailhouse, Xen, and Bao incur a negligible performance penalty, i.e., less than 1\% across all benchmarks. Although seL4 CAmkES-VMM also presents a small overhead for most benchmarks, the overhead can reach up to 7\%. 

For a virtualized system configured with a single guest VM, there are two main possible sources of overhead. The first source is the increase in TLB miss penalty due to the second stage of translation, since it can, in the worst case, increase the number of memory accesses in a page-walk by a factor of four. Second, the overhead of trapping to the hypervisor and performing interrupt injection, e.g., timer tick interrupt. Additionally, the pollution of caches and TLBs by the hypervisor might also affect guest performance.
To further understand the behavior of the benchmarks, in particular the larger overhead of the CAmkES-VMM, we have collected a number of microarchitectural events. Fig. \ref{fig:mibench-events} shows them normalized to the number of executed instructions. We highlight two events whose increase is highly correlated with the degradation observed: hypervisor L2 cache refills (Fig. \ref{fig:mibench-hyp-l2-miss}) and guest TLB misses (Fig. \ref{fig:mibench-guest-itlb-miss}), with Pearson correlation coefficients of up to 0.94 and 0.96, respectively.

\begin{figure*}[t]
    \centering
    \includegraphics[width=0.90\textwidth]{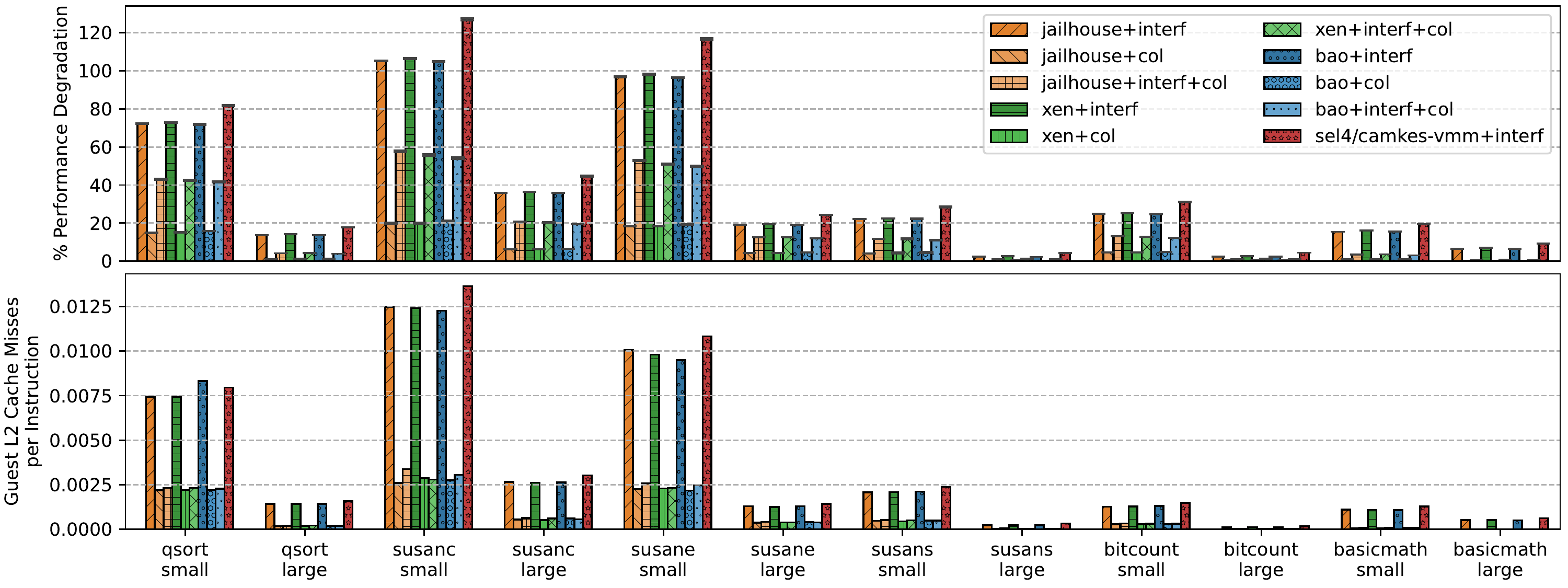}
    \vspace{-0.10cm}
    \caption{Performance degradation and L2 cache misses per instruction for the Mibench AICS under interference and coloring.}
    \label{fig:mibench-interf}
    \vspace{-0.45cm}
\end{figure*}


An important hypervisor feature to minimize the impact of two-stage translation is to leverage superpages. By inspecting hypervisor code, we concluded that only CAmkES-VMM does not have support for 2MiB superpages. This justifies the higher number of TLB misses. Notwithstanding, to corroborate this argument, we have configured the other SPH to preclude the use of superpages. As expected, we observed an increase in the performance degradation (and TLB misses) similar to CAmkES-VMM (Fig. \ref{fig:mibench-nosuper}). We still observed a gap of up to 2\% between CAmkES-VMM and the other SPH; this is related to the aforementioned interrupt handling and injection overheads, i.e., a consequence of the microkernel design: more costly switches between VM and VMM and a high number of VMM to microkernel calls for managing and inject the interrupts. This is confirmed by Figures \ref{fig:mibench-inst-ratio} and \ref{fig:mibench-hyp-excp}, which show the hypervisor-to-guest executed instruction ratio and the number of exceptions taken by the hypervisor, respectively. For these events, seL4 has a higher ratio when compared to the other SPH. We further investigate interrupt injection in Section \ref{sec:irq-latency}.

\begin{mdframed}[style=remarkstyle]

\mypara{Takeaway \thetakeawyacount \stepcounter{takeawyacount}.} SPH do not incur in meaningful performance impacts due to: (i) modern hardware virtualization support; (ii) 1-to-1 mapping between virtual and physical CPUs; and (iii) minimal traps. However, one key aspect is that SPH must have support for / make use of superpages to minimize TLB misses and page-table walk overheads.
\end{mdframed}

\mypara{Performance under interference.} We also evaluate inter-VM interference and the effectiveness of cache coloring at both guest and hypervisor levels. Fig. \ref{fig:mibench-interf} plots the results under interference (\textit{+interf}), with coloring enabled (\textit{+col}), and with interference and coloring enabled (\textit{+interf+col}). seL4 CAmkES VMM shows no results for coloring enabled as this feature is not openly available yet.

There are four conclusions to be drawn. Firstly, interference significantly affects the benchmark execution over all hypervisors. As expected, this is explained by a significant increase in L2 cache misses. On Jailhouse, Xen, and Bao performance is degraded by a similar factor, i.e., to a maximum of about 105\%; seL4-VMM is more susceptible to interference, reaching up to 125\% in the worst case. This pertains to the fact that, given that seL4-VMM executes a much higher number of instructions, the interference also impacts the execution of the hypervisor. Secondly, coloring, per se, significantly  impacts performance (up to about 20\%). This seems logical given that coloring (i) forces the hypervisor to use 4KiB pages, reducing TLB reach, and (ii) reduces the available cache space, which for working sets larger than LLC increases memory system pressure (i.e., L2 cache misses). Thirdly, coloring can only reduce interference but not completely mitigate it. 
In these experiments, the interference workload runs continuously. However, in a more realistic scenario, it might be intermittent. The improvement in predictability achieved by coloring is reflected in the difference between the base experiment results (bars in Fig. \ref{fig:mibench} and \textit{+interf} in Fig. \ref{fig:mibench-interf}) and respective variants with coloring enabled (\textit{+col} in Fig. \ref{fig:mibench-interf}). The lower the difference, the higher the predictability. For example, in the case of \textit{susanc-small}, we observed that without coloring, the variation can go up to 105 percentage points (pp), while when coloring is enabled, the observed overhead is around 58\%, which corresponds to a variation of 38 pp compared to the configuration with coloring enabled but without interference. Nevertheless, we observed that cache misses are essentially reduced to the same level as when coloring is enabled but without interference. Clearly, the observed interference is not only due to cache-line contention. There are points of contention at deeper levels of the memory hierarchy, e.g., buses and memory controller \cite{moscibroda2007} or even in internal LLC structures \cite{Valsan2016}. Finally, results on Xen and Bao demonstrate that hypervisor coloring has no substantial benefit as it only reduces performance degradation due to interference by at most 1\%  (omitted due to lack of space).

\begin{mdframed}[style=remarkstyle]

\mypara{Takeaway \thetakeawyacount \stepcounter{takeawyacount}.} Multicore memory hierarchy interference significantly affects guests' performance. Cache partitioning via page coloring is not a silver bullet as despite fully eliminating inter-core conflict misses, it does not fully mitigate interference (up to 38 pp increase in relative overhead).
\end{mdframed}
\vspace{-0.25cm}

\section{SPH: Interrupt Latency} \label{sec:irq-latency} 

As discussed in Section \ref{sec:arm-virt}, the existing GIC virtualization support is not ideal for MCS: hypervisors have to handle and inject all interrupts and must actively manage list registers when the number of pending interrupts is larger than the physical list registers. This is of particular importance to guarantee the correct interrupt priority order which might be critical for an RTOS \cite{Hofer2009}.
In this section, we investigate the overhead of each SPH in the interrupt latency, their susceptibility to interference, and the effectiveness of cache coloring. Then, we evaluate the direct injection technique and analyze interrupt priority support as well as virtual IPI latencies.

\mypara{Methodology.} To measure interrupt latency, we used a custom lightweight baremetal benchmark, which measures the latency of a periodic interrupt triggered by the Arm Generic Timer. The timer is programmed in auto-reload mode, to continuously trigger an interrupt at each 10 ms. The interrupt handler reads the value of the timer, i.e., it measures the time elapsed since the interrupt was triggered. Each measurement is carried out with cold L1 caches. To achieve this, after each measurement, we flush the instruction cache. During the 10 ms, we also prime the L1 data cache with useless data.

\begin{figure}
    \centering
    \includegraphics[width=0.38\textwidth]{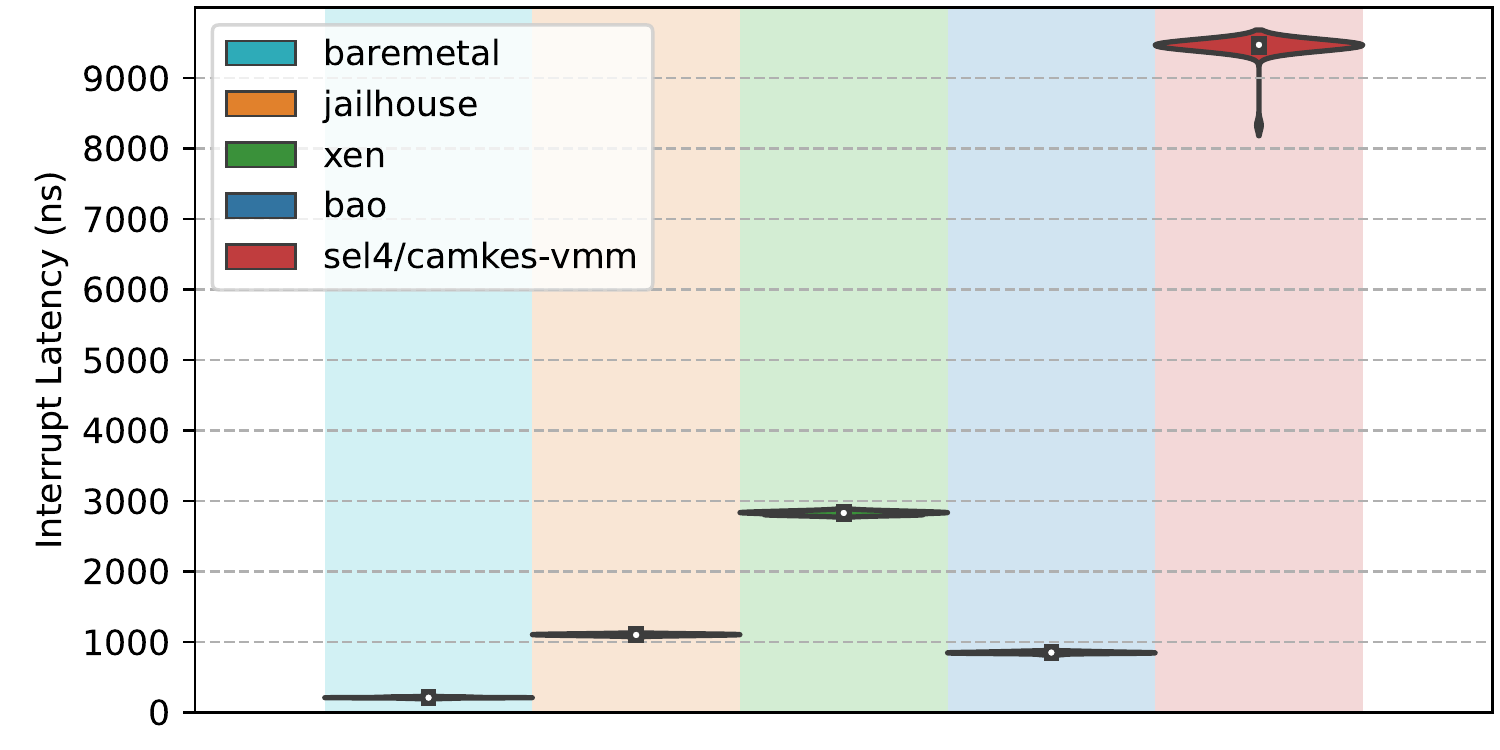}
    \caption{Base interrupt latency.}
    \label{fig:irqlat-base}
    \vspace{-0.6cm}
\end{figure}

\mypara{Base Latency.} Fig. \ref{fig:irqlat-base} depicts the violin plots for the custom benchmark running atop each SPH. From the baseline of about 200 ns, Bao and Jailhouse incur the smallest increase, albeit significant, to an interrupt latency of about 4x (840 ns) and 5x (1090ns), respectively. Xen shows an increase of about 14x (2800 ns). The variance observed in these three systems is negligible. The difference observed between Jailhouse/Bao and Xen is justified by the interrupt injection path being highly optimized in the former, while more generic in Xen. We confirmed this by studying the source code and assessing the number of instructions executed by each hypervisor on the interrupt handling and injection path: while Jailhouse and Bao execute around 200 instructions, Xen executes about 1050.


seL4-VMM presents the largest interrupt latency (47x, 9400 ns), an order of magnitude higher than Jailhouse and Bao. The variance of the latency is also affected. This can be explained by the interrupt handling and injection mechanism of a microkernel architecture. In the other SPH, each interrupt results in a single exception taken at EL2, where the interrupt is handled and injected in the VM; virtualization support is leveraged such that no further traps occur. In CAmkES VMM it results in four traps to the microkernel: (i) the first due to the interrupt that results in forwarding it as a message to the VMM; 
(ii) a system call from the VMM to inject the interrupt in the VM (i.e., write the list register); (iii) another to ``reply" to the exception, resuming the VM; and (iv) a final  one where the VMM waits for a message signaling a new VM event or interrupt, resulting in a final context-switch back to the VM. We have also concluded that seL4 does not use a GIC feature that would allow guests to directly deactivate\footnote{Deactivating an interrupt in the GIC means marking it as handled, enabling the distributor to forward it to the CPU when it occurs again.} the physical interrupt, resulting in an extra trap.

\begin{mdframed}[style=remarkstyle]




\mypara{Takeaway \thetakeawyacount \stepcounter{takeawyacount}.} Due to the lack of efficient hardware support for directly delivering interrupts to guests in Arm platforms, all SPH increase the interrupt latency by at least one order of magnitude. However, by-design, SPH such as Jailhouse and Bao are able to achieve the lowest latencies as they provide an optimized path for hardware interrupt injection.
\end{mdframed}
\vspace{-0.2cm}

\begin{figure} [!t]
    \centering \hspace{-0.4cm}
    \includegraphics[width=0.485\textwidth]{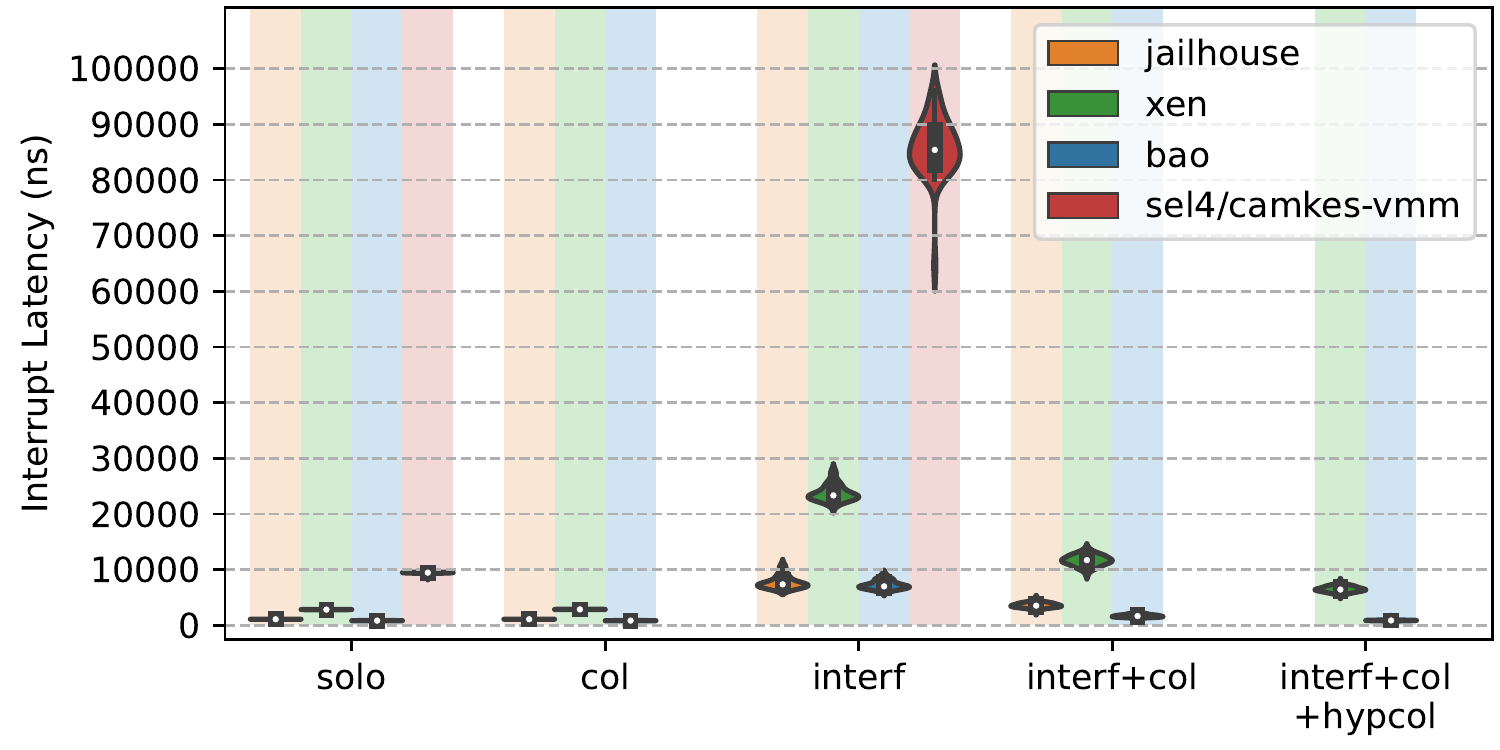}
    \hspace{0.4cm}
    \vspace{-0.7cm}
    \caption{Interrupt latency under interference and cache coloring.}
    \label{fig:irqlat-interf}
    \vspace{-0.25cm}
\end{figure}

\mypara{Latency Under Interference.} Fig. \ref{fig:irqlat-interf} shows the results for interrupt latency under interference, including the baseline results of Fig. \ref{fig:irqlat-base} for relative comparison as \textit{solo}. Analyzing the effects of VM interference on interrupt latency (\textit{interf}), we observed that Bao latency increases to an average of 7260 ns, Jailhouse to 7730 ns, Xen to 23000 ns, and seL4-VMM to 85940 ns. It corresponds to an increase of 36x, 38x, 115x, and 430x, respectively, compared to the base latency. It is also worth noting that the variance also increases. When enabling coloring (\textit{col}), we measured no significant difference in interrupt latency compared to the base case. However, when enabling cache coloring in the presence of inter-VM interference (\textit{interf+col}), there is a visible improvement in average latency and variance. However, note that the observed variance does not constitute a measure of predictability. As explained in Section \ref{sec:perf}, predictability is reflected in the difference between the \textit{interf} and \textit{interf+col} results and respective baselines, i.e., \textit{solo} and \textit{col}. Finally, by applying coloring also to the hypervisor (\textit{interf+col+hypcol}), Bao latency is reduced to almost no-interference levels with negligible variance. Xen latency also drops considerably to an average of 6300 ns.

\begin{figure} [!t]
    \centering
    \begin{subfigure}{0.5\textwidth}
        \centering
        \begin{subfigure}{0.475\textwidth}
            \centering
            \includegraphics[width=1\textwidth]{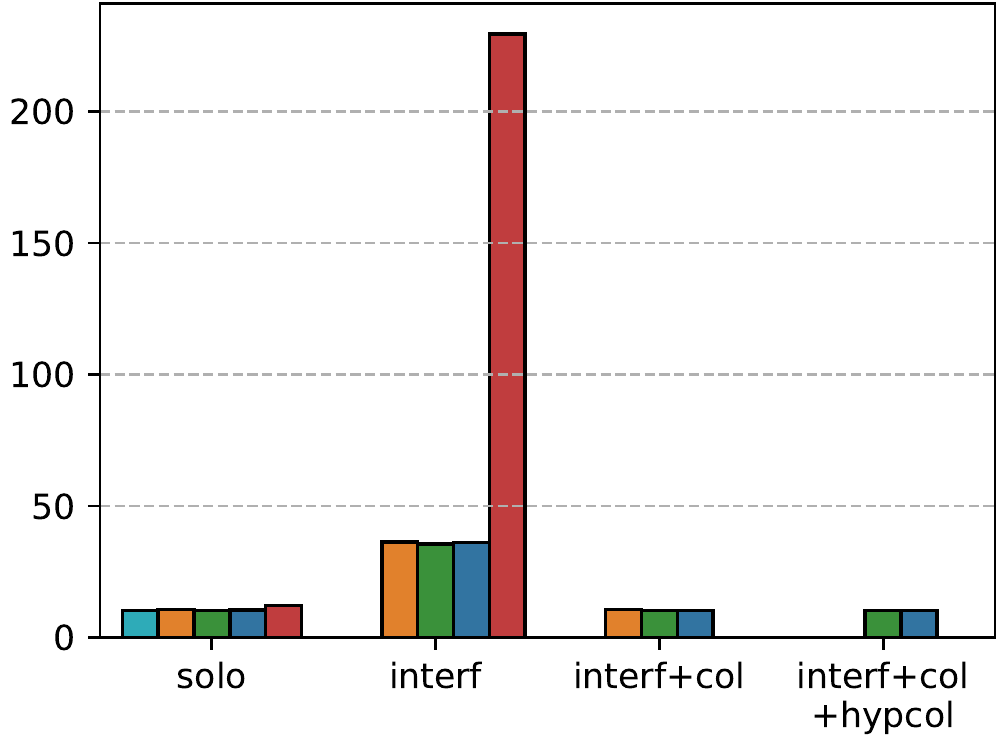}
            \vspace{-0.6cm}
            \caption{Guest L2 cache misses.}
            \label{fig:irqlat-l2miss-el1}
    \end{subfigure}
    \begin{subfigure}{0.475\textwidth}
        \centering
        \includegraphics[width=1\textwidth]{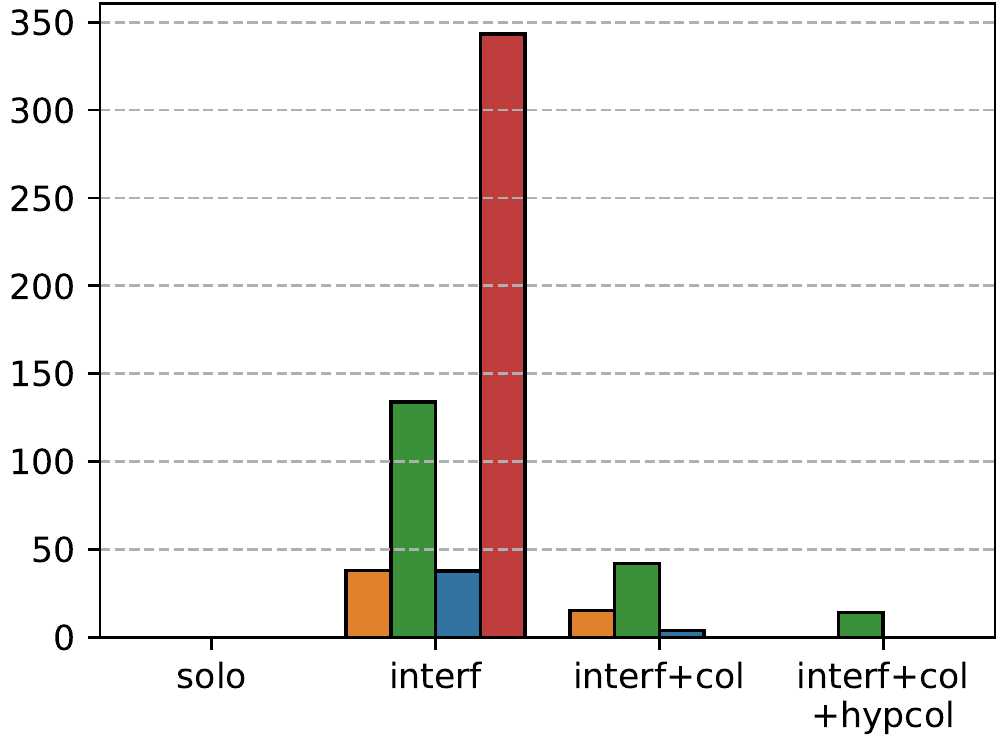}
        \vspace{-0.6cm}
        \caption{Hypervisor L2 cache misses.}
        \label{fig:irqlat-l2miss-el2}
    \end{subfigure}
    \vspace{0cm}
    \end{subfigure}
    \caption{L2 cache misses for the interrupt latency benchmark.}
    \label{fig:irqlat-l2miss}
    \vspace{-0.6cm}
\end{figure}

The observed interrupt latency under interference can be mostly explained by L2 cache misses. Fig. \ref{fig:irqlat-l2miss} shows the L2 cache misses for both guest and hypervisor during interrupt latency measurement. We can see that interference increases guest L2 cache misses, but that cache coloring can lower them back to the base case values. However, this is not the case for hypervisor L2 cache misses. For the base case, there are no cache misses for the hypervisor, which increases substantially under interference. Despite VM coloring contributing to reduce hypervisor L2 cache misses, only by coloring the hypervisor level, it is possible to minimize L2 cache misses for the hypervisor. On Bao, L2 cache misses are fully eliminated, but not on Xen\footnote{At the time of writing, Xen's coloring patch was still under review. Thus, the assessed implementation may contain some imprecisions that are likely to be fixed by the time the patch is merged.}, which might explain why latency does not reduce to non-interference levels.

\begin{mdframed}[style=remarkstyle]



\mypara{Takeaway \thetakeawyacount \stepcounter{takeawyacount}.} Interrupt latency increases tenfold under the interference workload. Applying cache coloring to VMs proves very beneficial, but for it to be fully effective, it is imperative to reserve a color for the hypervisor itself.
\end{mdframed}
\vspace{-0.2cm}

\begin{figure} [!t]
    \centering
    \includegraphics[width=0.4\textwidth]{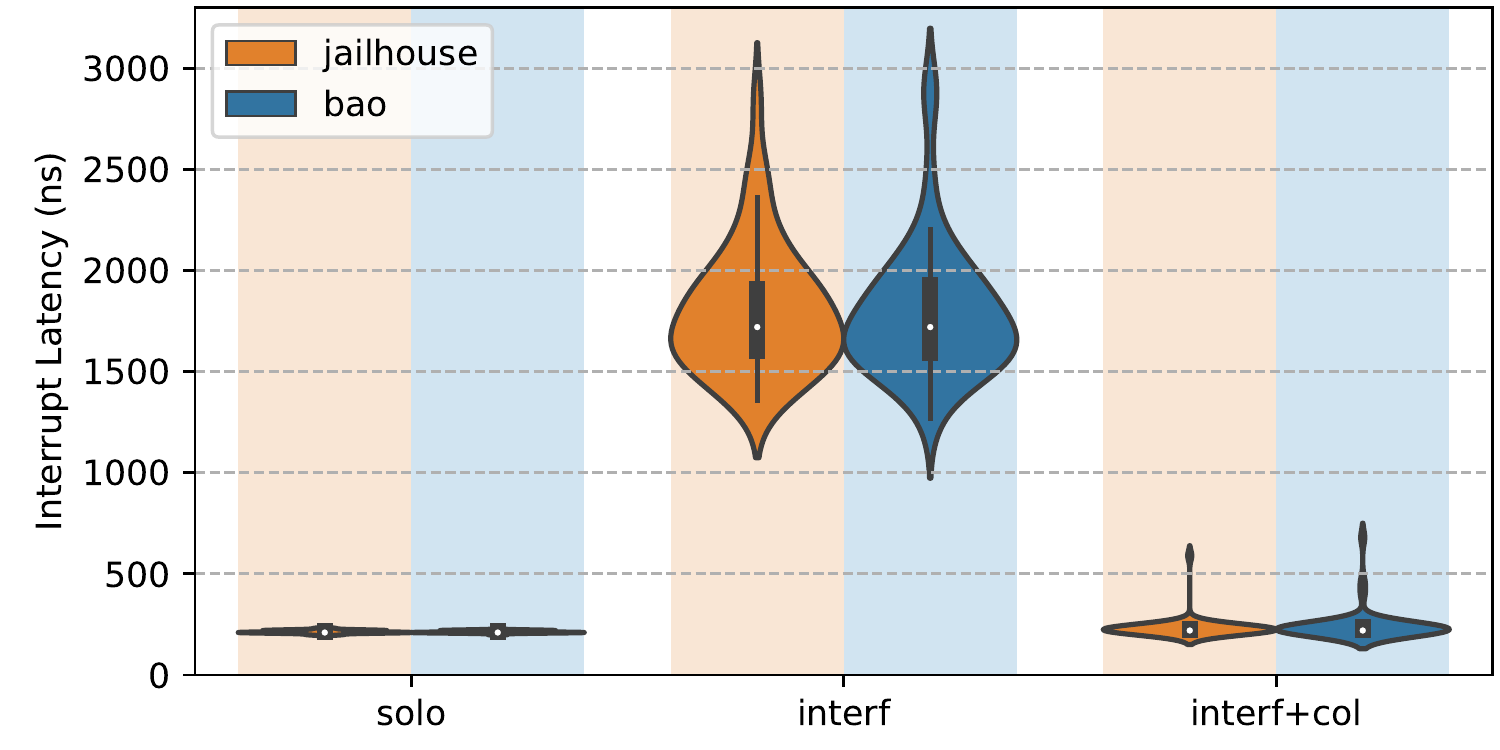}
    \vspace{-0.2cm}
    \caption{Interrupt latency with direct injection enabled.}
    \label{fig:irqlat-di}
    \vspace{-0.3cm}
\end{figure}

\mypara{Direct Injection.} We evaluate the effectiveness of the direct injection technique, implemented only in Jailhouse and Bao. Fig. \ref{fig:irqlat-di} depicts the results. The first conclusion is that for the base case, i.e., no interference, the interrupt latency is near to native (about 210 ns). Indeed, we have confirmed that during the execution of the benchmark, there are no traps to the hypervisor. Next, we observed that interference somewhat increases latency and its variance, but much less than in the previous experiments. Finally, we concluded that by enabling coloring, it is possible to lower the average latency to near native (243 and 232 ns for Bao and Jailhouse, respectively), however, there is still some variance due to the interference.

\begin{mdframed}[style=remarkstyle]

\mypara{Takeaway \thetakeawyacount \stepcounter{takeawyacount}.} The direct injection technique is effective in addressing the shortcomings of GIC interrupt virtualization, as results clearly demonstrated interrupt latency overhead is reduced to near native latencies.




\end{mdframed}
\vspace{-0.2cm}

\begin{figure}
    \centering
    \includegraphics[width=0.4\textwidth]{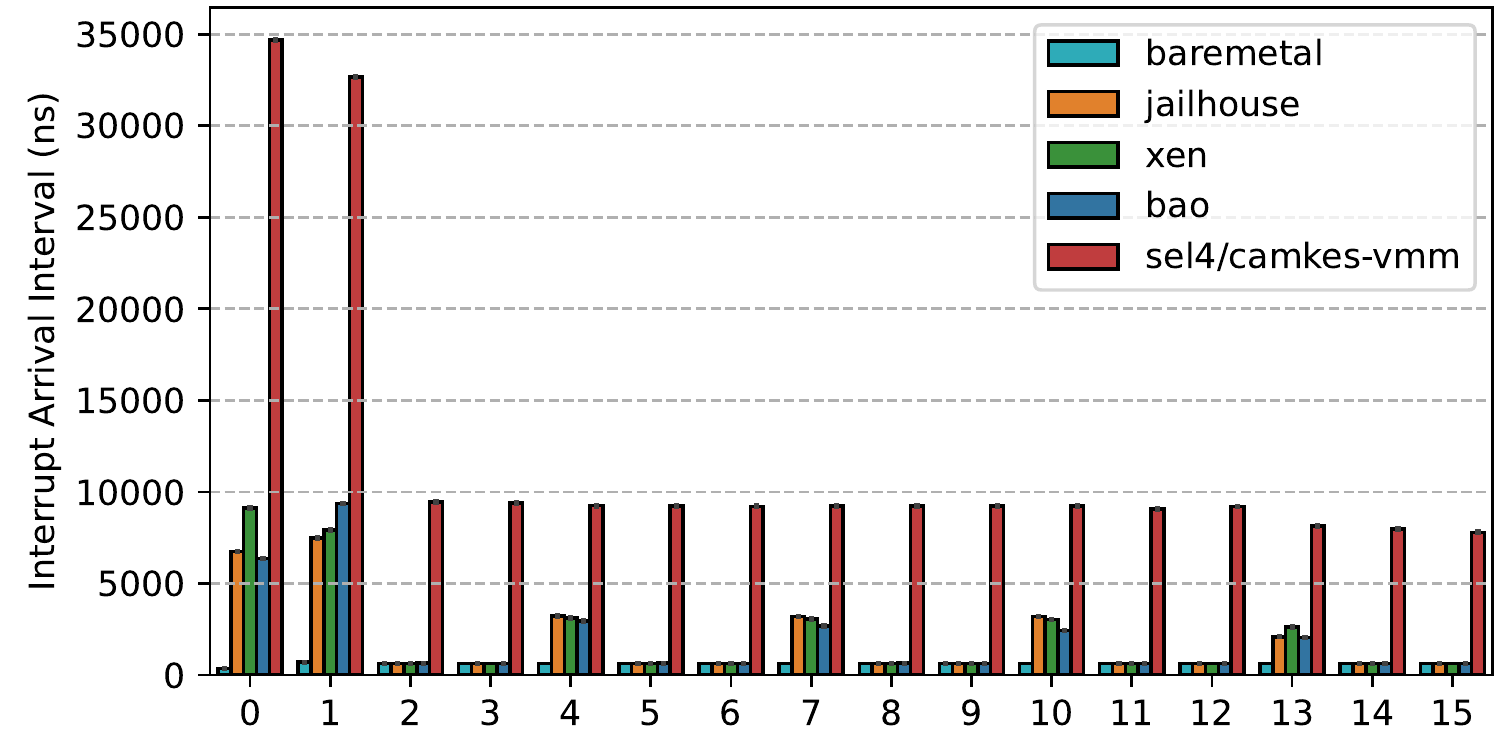}
    \vspace{-0.2cm}
    \caption{Time between the handling of different priority interrupts triggered simultaneously, i.e., for interrupt N in the X-axis, the time between the arrival of interrupts N-1 and N.}
    \label{fig:irqprio}
    \vspace{-0.55cm}
\end{figure}

\mypara{Priority Handling.} For studying the support of SPH delivering interrupts in the correct priority order, we have implemented a PL device which can be used to trigger up to 16 simultaneous interrupts, and a custom benchmark that assigns each a different priority. It starts by triggering the eight lowest priority interrupts. When handling the first, it triggers the eight highest priority interrupts. This would force the hypervisor to refill the four available LRs with the new higher priority interrupts, and refill them in priority order as LRs become available. The benchmark verifies if the priority order was kept and measures the arrival interval between each interrupt.
We verified that only Xen and Bao guarantee the delivery of interrupts in the correct priority order.
By inspecting the code, we have confirmed that both seL4-VMM and Jailhouse fill the GIC LRs following a FIFO policy. Furthermore, the seL4-VMM does not even commit the interrupt priorities configured in the virtual GIC distributor to hardware, precluding the arrival of physical interrupts in the correct priority order. Fig. \ref{fig:irqprio} shows that across all hypervisors if multiple interrupts are delivered simultaneously, there is an increase by several orders of magnitude in the arrival time of the first and second interrupts, which is less than 700 ns for the baremetal case. This larger increase is justified by the fact that the hypervisors must handle all interrupts before the guest starts handling the first interrupt. Another observation is that there is a periodic increase in the interval of arrival. We have concluded this is the point at which there are no pending interrupts left in the LRs, which triggers the hypervisor to refill these registers with previously spilled pending interrupts.

\begin{mdframed}[style=remarkstyle]

\mypara{Takeaway \thetakeawyacount \stepcounter{takeawyacount}.} Only Xen and Bao respect interrupt priority order. Additionally, we observe that for all SPH, if multiple interrupts are triggered simultaneously, there is a partial priority inversion as lower priority interrupts take precedence due to the need for the hypervisor to handle and inject them.




\end{mdframed}
\vspace{-0.2cm}

\mypara{Inter-Processor Interrupts.} IPIs (SGIs) are critical for multicore VM performance. For a vCPU to send an SGI, the guest must write a virtual GIC distributor register. This will trap to the hypervisor that must emulate the access and forward the event to the target core, where the SGI is injected via list registers. We use a custom baremetal benchmark to measure IPI latency. It works by measuring the time between when the source vCPU writes the distributor register and when the final IPI handler starts executing. It also measures the overhead of the trap. We instrument the SPH to sample the time the IPI is forwarded internally; this signals the end of the emulation and translates the overhead of injecting the interrupt in the target.

\begin{figure}
    \centering
    \includegraphics[width=0.44\textwidth]{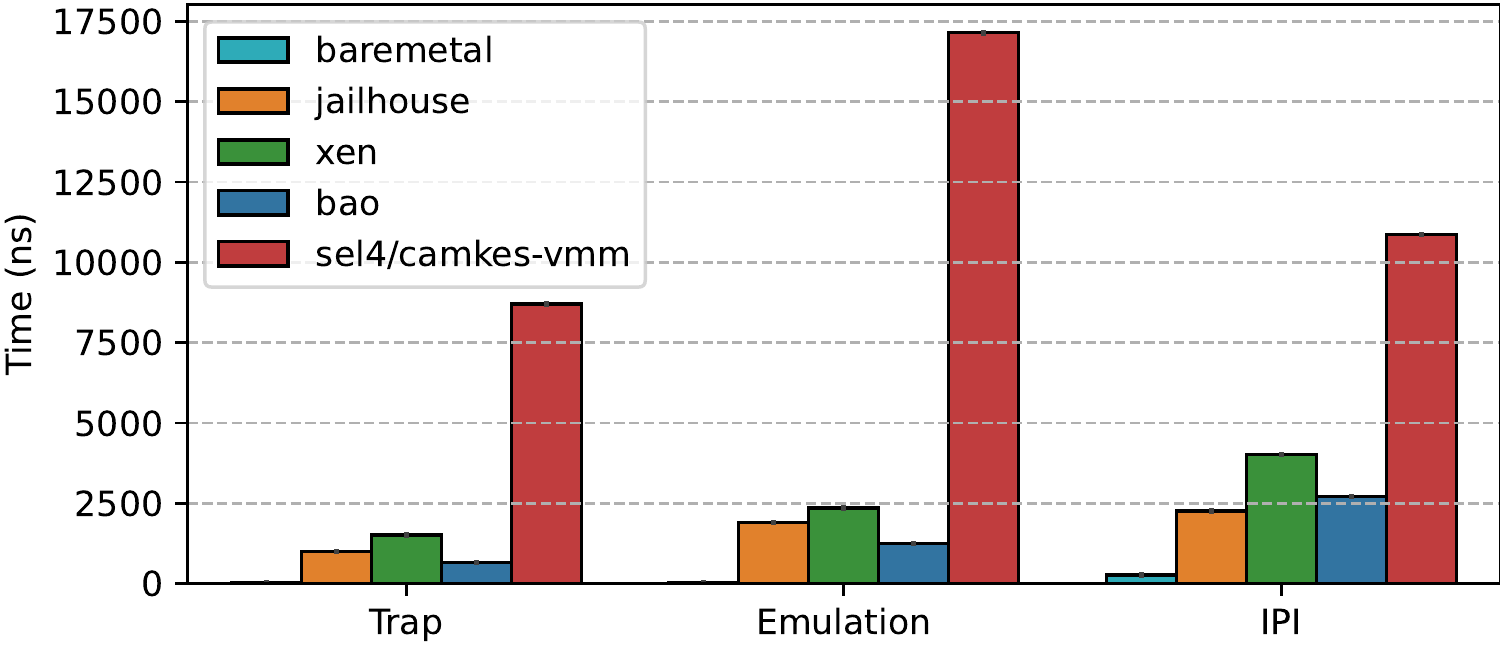}
    \caption{Average cost for each send IPI operation component.}
    \label{fig:ipi}
    \vspace{-0.5cm}
\end{figure}

Figure \ref{fig:ipi} shows that IPI latency increases significantly for all SPH. While the baremetal IPI latency is around 260 ns, it reaches 2258 ns for Jailhouse, 4157 ns for Xen, 2711 ns for Bao, and 10868 ns for the CAmkES VMM. However, the costs of the register access emulation and interrupt injection are not proportional across all SPH. For example, Bao has the lowest emulation and event forwarding times, but the overall IPI latency is higher than Jailhouse's. This means that the interrupt injection path on Bao is slower than on Jailhouse. By inspecting the source of both hypervisors, we have observed that Bao immediately forwards the SGI event to the target core, performing all interrupt injection operations in the target core. Jailhouse, in turn, manages the interrupt injection structures at the source core and only then signals the target vCPU by writing the list register. Xen follows the same approach as Jailhouse, but presents higher overhead. The CAmkES VMM has the highest overhead due to the large number of system calls the VMM issues to the microkernel (in total, 7). Four are issued before the event forwarding, and the rest only after the SGI is forward to the target core. All in all, the access to the virtual distributor is more expensive than the IPI itself.

\begin{mdframed}[style=remarkstyle]
\mypara{Takeaway \thetakeawyacount \stepcounter{takeawyacount}.} IPI latency reflects the same overheads of external interrupts. Future Arm platforms might reduce them with GICv4.1 \cite{arm2022gic}. In the short term, direct injection might alleviate this issue. However, both approaches fall short of achieving native latency as they still pay the price of emulating the write to the ``IPI send" register.
\end{mdframed}



\section{SPH: Inter-VM communication} \label{sec:vm-comm}
For inter-VM communication, SPH typically only provide statically allocated shared memory. This is usually coupled with an asynchronous notification mechanism signaled as an interrupt. All four SPH provide such mechanisms. Next, we analyze inter-VM notification latency and transfer throughput.

\begin{figure} [!t]
    \centering
    \includegraphics[width=0.45\textwidth]{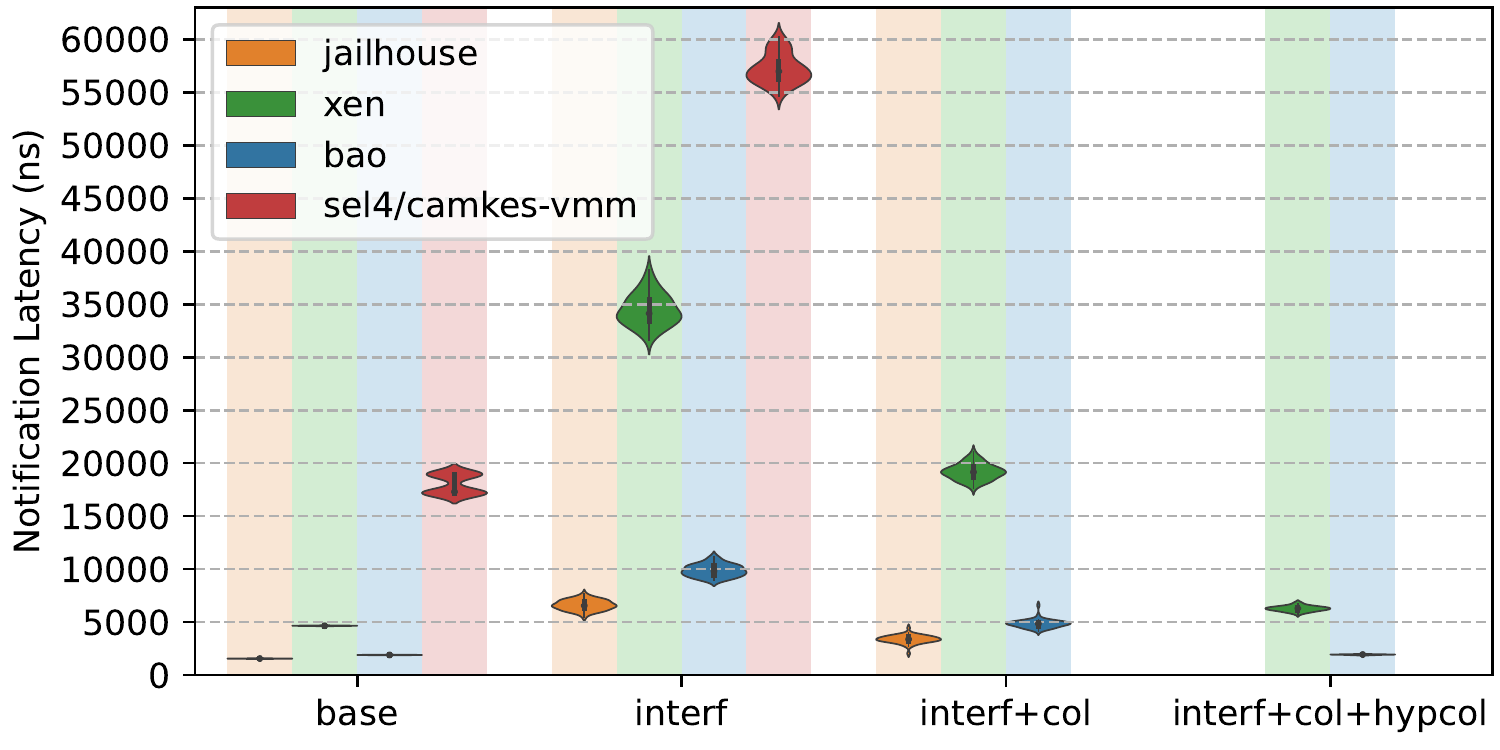}
    \caption{Inter-VM notification latencies.}
    \label{fig:notification}
    \vspace{-0.5cm}
\end{figure}

\mypara{Inter-VM latency.} Fig. \ref{fig:notification} shows the inter-VM notification latency, reflecting the time since the notification is issued until the execution of the handler in the destination VM. The relative differences between the latencies for each SPH are similar to those observed for passthrough interrupts and IPIs. Jailhouse achieves the lower latency (1500 ns), followed by Bao (1900 ns). Xen shows an intermediate value of 4600 ns, while seL4 CAmkES VMM is significantly larger than others (average 18000 ns). Studying the internals of the implementations, we note that while most hypervisors synthesize and inject the virtual interrupts, Jailhouse uses non-allocated physical interrupts for these notifications. Thus, to send one, Jailhouse only sets the interrupt pending in the GIC distributor. This is significantly advantageous when combined with direct injection. Note that enabling direct injection in Bao would preclude the use of this mechanism. For seL4, we highlight the impact of the microkernel architecture since atop VM/VMM context switches, we observe additional overheads due to inter-VMM communication. Lastly, we see that interference increases all latencies accordingly and that coloring can mitigate it.

\begin{figure}[!t]
    \centering
    \includegraphics[width=0.48\textwidth]{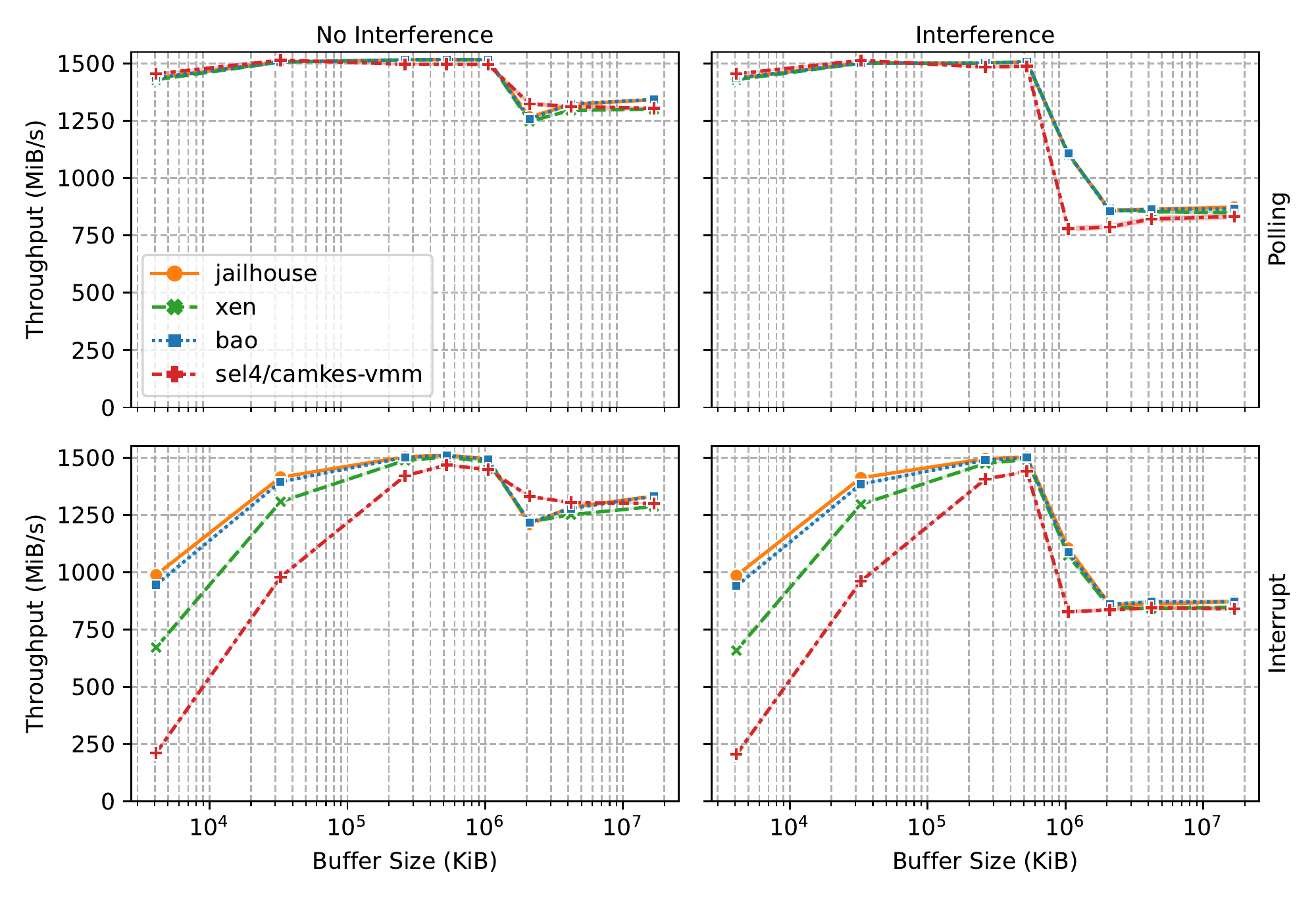}
    \vspace{-0.2cm}
    \caption{Inter-VM communication throughput.}
    \label{fig:comm}
    \vspace{-0.5cm}
\end{figure}

\mypara{Inter-VM throughput.} In Fig. \ref{fig:comm}, we evaluate the throughput of bulk data transfers via a shared memory buffer. The benchmark transmits 16 MiB of random data through a shared buffer with varying sizes. When the source VM finishes writing the buffer, it either signals the destination VM via a shared memory flag or via an asynchronous notification, and waits for a signal back to start writing the next chunk. For the polling scenario, the obtained throughput is very similar across all hypervisors; this confirms that are no significant differences in how they allocate and map memory or configure memory attributes. Throughput is stable (1500 MiB/s) until the buffer size surpasses the LLC size (1 MiB), dropping to about 1300 MiB/s. For the asynchronous scenario, throughput is significantly impacted when using smaller buffer sizes, given the high number of synchronization points that reflects the observed interrupt overheads. Finally, we note that interference has no significant effect as long as the buffer size is kept below about half the size of LLC. Beyond that, throughput is reduced from 1300 to 850 MiB/s. Although not shown due to lack of space, using coloring does not prove beneficial, as the throughput illustrated in Fig. \ref{fig:comm} remains virtually unchanged.

\begin{mdframed}[style=remarkstyle]
\mypara{Takeaway \thetakeawyacount \stepcounter{takeawyacount}.} Inter-VM notification latencies are significant and, as is the case for hardware interrupts, very susceptible to the effects of interference. However, for bulk data transfers it does not seem to significantly affect throughput if the shared buffer size is chosen on a range of about one-fourth to half the LLC size (i.e., 256 KiB to 512 KiB).
\end{mdframed}
\vspace{-0.335cm}

\section{SPH: Boot time} \label{sec:boot}
System boot time is a crucial metric in industries such as automotive \cite{Hamelin2020, Golchin2022} as critical components have strict timing requirements for becoming fully operational. 
\mypara{Platform's Boot Flow.} The platform's boot flow \cite{xilinxtrm} starts by executing ROM code which loads the first-stage bootloader (FSBL) and enables the main cores. These initial boot stages setup the platform basic infrastructure (e.g., clocks, DRAM) and load the TF-A and U-boot. U-boot will load the hypervisor and, except for Jailhouse, the guest images. 
Bao and Xen directly boot guests after initialization. Jailhouse starts with the boot of the Linux root cell, that installs the hypervisor which then loads the guests. seL4's execution starts with an ELF loader which loads the all images, initializes secondary cores, and sets up an initial set of page tables for the microkernel. The microkernel initializes and hands control to user space.

\begin{figure}[!t]
    \centering
    \includegraphics[width=0.465\textwidth]{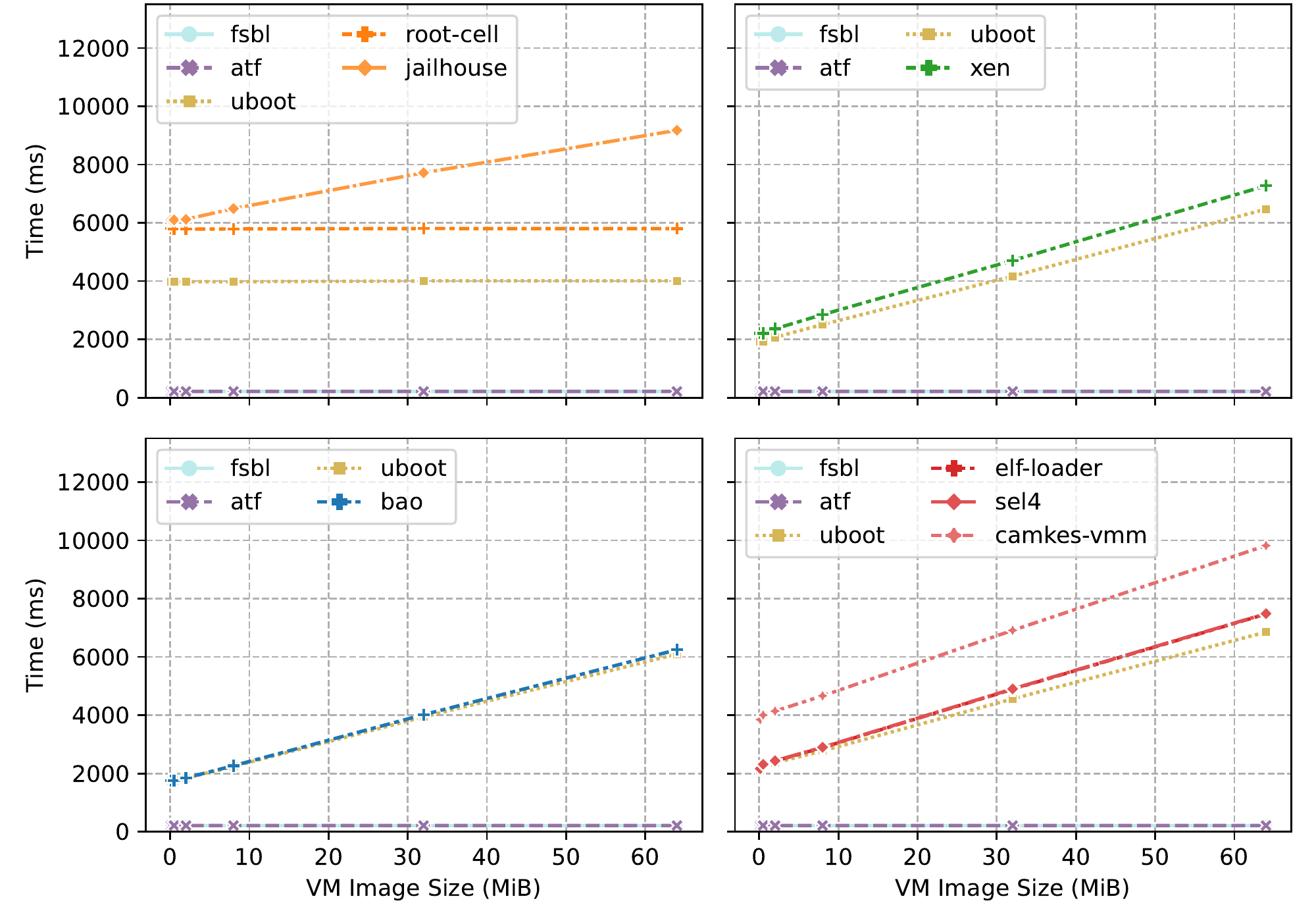}
    \vspace{-0.15cm}
    \caption{Boot time for each stage by VM image region size.}
    \label{fig:boottime-cumulative}
    \vspace{-0.5cm}
\end{figure}

\mypara{Total VM Boot Time.} The hypervisor boot time is heavily dependent on the VM and how it is configured. We observed that the VM image size is one of the parameters that has the higher impact in the hypervisor boot time. We measure boot time as a function of VM image size. Thus, to understand the overhead of the hypervisor in the context of the complete boot flow, in Fig. \ref{fig:boottime-cumulative}, we plot the cumulative time for each boot stage. Here, we can confirm that in all hypervisors but Jailhouse, the bulk of boot time is spent by U-boot. For Jailhouse, U-boot run time is constant, albeit large, as it always only loads the root cell's image. Jailhouse execution time increases steeply while loading the VM image. From this macro perspective, the other hypervisors add an almost constant offset to U-boot's boot time, the largest being seL4-VMM's. We observe this overhead is not on the microkernel, but at user level, which nevertheless heavily interacts with the microkernel to setup capabilities and kernel objects. We can conclude that VM boot time has its bottleneck by the loading of guest images to memory, not the hypervisor logic.

\mypara{FreeRTOS and Linux Boot Times.} We also measure the boot time of (i) a small VM running FreeRTOS with a 90 KiB image and (ii) a large VM with a Linux guest (built-in ramfs) totaling 59 MiB of image size. For Jailhouse, the Linux VM is a non-root cell. In Table \ref{tab:boottime-vms}, we present results for a single-guest and a dual-guest system. For the latter, both VMs boot simultaneously; thus, we did not run experiments for dual-guest with Jailhouse, because it launches VMs sequentially. 
Table \ref{tab:boottime-vms} presents the absolute boot time for the guest's native and virtualized execution, highlighting the relative percentage increase compared to native execution. For the single-guest FreeRTOS VM, all hypervisors but Bao cause a non-negligible increase in boot time. The same happens with the single-guest Linux VM. For the dual-guest configuration, we concluded that the small VM is heavily affected for all hypervisors. Surprisingly, we observe that although the cost of booting a single FreeRTOS in Bao is negligible, this is not true for a dual-guest configuration. Booting it alongside a Linux VM significantly increases its boot time, reaching similar overheads to those observed in Jailhouse's sequential boot.

\begin{mdframed}[style=remarkstyle]
\mypara{Takeaway \thetakeawyacount \stepcounter{takeawyacount}.} The major bottleneck for the VM boot time is caused by the bootloader, not the hypervisors. Notwithstanding, the hypervisor can significantly increase the boot time of a critical VM (small RTOS) when booting it alongside a larger VM (e.g., in dual-OS Linux+RTOS configuration).


\end{mdframed}
\vspace{-0.4cm}

\section{SPH: Code Size and TCB} \label{sec:code-tcb}

In MCS, the size of the hypervisor code, measured in source lines of code (SLoC), is critical. It should be minimal as it is part of the trusted computing base (TCB) of all VMs. 
In this paper, we consider that a VM TCB encompasses any component with sufficient privileges that if it is compromised or malfunctions, might be able to affect the safety and/or security properties of the VM. As well understood in the literature, a larger TCB typically has a higher number of bugs and wider attack surface \cite{Biggs2018}, resulting in a higher probability of vulnerabilities. It is important to understand that each VM has its own TCB. Thus, CAmkES VMM is only considered for the managed VM's TCB, not the others. Further, large code bases are impractical for certification, both from a technical and economic perspective. To qualify a component assigned a safety integrity level (SIL), all components on which it depends must also be qualified to the same or higher SIL \cite{Esper2018}.


\begin{table}[!t]
\resizebox{1\linewidth}{!}{
\begin{tabular}{lllllll}
\toprule
                          &        & Baremetal    & Jailhouse                & Xen               & Bao                       & seL4-VMM             \\ \midrule
\multirow{2}{*}{FreeRTOS} & Single & 1670.89     &  6242.18 / 173.58\%     & 2338.24 / 39.94\%   & 1716.23 / 2.71  \%  &  3496.19 / 109.24\%   \\ \cmidrule{2-7} 
                          & Dual   &             &  N/A                     & 6887.88 / 312.23\% & 5734.04 / 143.17\%   &  9291.02 / 456.05\%   \\ \midrule
\multirow{2}{*}{Linux}    & Single & 7665.14     &  12284.92 / 60.27\%     & 8533.88 / 11.33\%   & 7805.54 / 1.83  \%  &  12629.79 / 64.77\%   \\ \cmidrule{2-7} 
                          & Dual   &             &  N/A                     & 8707.15 / 13.59\%  & 7895.95 / 3.01  \%   &  13086.86 / 70.73\%   \\ \bottomrule
\end{tabular}
}
\caption{Total boot time (ms) and relative increase compared to the baremetal case, for FreeRTOS and Linux VMs.}
\label{tab:boottime-vms}
\vspace{-0.2cm}
\end{table}

\begin{table}[!t]
\centering
\resizebox{0.8\linewidth}{!}{
    \begin{tabular}{ccccclc}
        \toprule
        \multicolumn{2}{c}{}                                   & \multicolumn{2}{c}{C}     & \multirow{2}{*}{Asm} & \multicolumn{1}{c}{\multirow{2}{*}{\begin{tabular}[c]{@{}c@{}}Total\\ (SLoC)\end{tabular}}} & \multirow{2}{*}{\begin{tabular}[c]{@{}c@{}}\textit{.text} \\ (KiB)\end{tabular}} \\ \cmidrule{3-4}
                                                                &             & \multicolumn{1}{c}{.c}    & .h    &      & \multicolumn{1}{c}{} & \\ \midrule
        \multicolumn{1}{c}{\multirow{2}{*}{\textbf{jailhouse}}} & hypervisor  & \multicolumn{1}{c}{7308}  & 2279  & 342  & 9929  & 79.3  \\ \cmidrule{2-7} 
        \multicolumn{1}{c}{}                                    & driver      & \multicolumn{1}{c}{2041}  & 139   & N/A  & 2180  & 20.1  \\ \midrule
        \multicolumn{2}{c}{\textbf{xen}}                                      & \multicolumn{1}{c}{57360} & 8127  & 1765 & 67342 & 451.5 \\ \midrule
        \multicolumn{2}{c}{\textbf{bao}}                                      & \multicolumn{1}{c}{5046}  & 2840  & 537  & 8423  & 57.9  \\ \midrule
        \multicolumn{1}{c}{\multirow{2}{*}{\begin{tabular}[c]{@{}c@{}}\textbf{seL4}\\\textbf{CAmkES VMM}\end{tabular}}}      & microkernel & \multicolumn{1}{c}{14569} & N/A   & 189  & 14758  & 224.7 \\ \cmidrule{2-7} 
        \multicolumn{1}{c}{}                                    & VMM         & \multicolumn{1}{c}{20932} & 19291  & N/A  & 40223 & 724.3 \\ \bottomrule
        \end{tabular}
}
    \caption{Hypervisor SLoC count and binary code size.}
    \label{tab:tcb-size}
    \vspace{-0.6cm}
\end{table}

\mypara{Methodology.} We measured SLoC for the target configurations using \textit{cloc} \cite{cloc}. Xen build system offers a make target to assess the SLoC for a specific configuration. However, it does not count header files, which we believe must be accounted for since they provide function-like macros and inline functions. We have modified the Xen makefile to measure headers. We have also extended Jailhouse and Bao build systems with the same functionality. For seL4, we used the fully unified and pre-processed kernel source file to assess the microkernel code base. For the CAmkES VMM, given that its source code is scattered throughout multiple seL4 project libraries, we were not able to list its source code files from the build system. Instead, we used debug information from the final executable and inspected each source to assess the included header files. 


\mypara{Code Size.} Looking at Table \ref{tab:tcb-size} we see Bao and Jailhouse have the smallest code base of about 8400 and 9900 SLoC, respectively. Bao is implemented as a standalone component with no external dependencies. However, since part of Jailhouse functionality is implemented as a Linux kernel module, we also account that for the code base. It adds about 2180 SLoC, bringing Jailhouse total code base to 12 KSLoC. For Xen we use a custom config with almost all features disabled, except a few ones such as coloring and static shared memory. It features the largest code base with around 67 KSLoC. Finally, seL4 microkernel has 14.5 KSLoC, while the CAmkES VMM can go up to 40K, i.e., almost 55 KSLoC in total. The visible difference between Bao and Jailhouse, and seL4 microkernel and, especially, Xen, lies in the fact that the former were designed specifically for the static partitioning use case, while the latter aim at being more generic and adaptable. These differences are reflected in the binary size of each hypervisor.

\mypara{TCB.} The hypervisor SLoC does not directly reflect the VM TCB. Although by design SPH such as Bao has a smaller SLoC count, the seL4-VMM is vastly superior from a security perspective: shared TCB is limited only to the formally verified microkernel, because each VM is managed by a fully isolated VMM. From a FuSa certification standpoint, however, the VMM would still need to be considered. Moreover, seL4 formal proofs are limited to a set of kernel configurations, currently not including multicore. Regarding Jailhouse, despite its small size, the root cell is a privileged component of the system. It executes part of all VM management logic, being in the critical path for booting all other VMs. It is arguably part of all VM's TCB, increasing it significantly \cite{Biggs2018}.
Analogously, Xen must depart from true Dom0-less to leverage richer features (e.g., PV drivers, dynamic VM creation). Recently, the Xen community has ignited efforts to use a smaller OS, such as Zephyr \cite{zephyr_project_2023}, as Dom0, refactor Xen to MISRA C, and provide extensive requirements and test documentation \cite{Mygaiev2021}.

\begin{mdframed}[style=remarkstyle]
\mypara{Takeaway \thetakeawyacount \stepcounter{takeawyacount}.} Hypervisors specifically targeting static partitioning have the smallest code bases. Despite facilitating certification, none of the evaluated SPH provide other artifacts (e.g., requirements specification, coding standards). Xen is the first to take steps in this direction; nevertheless, seL4's formal proofs provide the most comprehensive guarantees.
\end{mdframed}
\vspace{-0.45cm}

\section{Discussion and Future Directions} \label{sec:discussion} 
In this section, we discuss some of the open issues and potential research directions to improve the guarantees of SPH.

\mypara{Interference Mitigation Techniques.} Cache coloring does not fully mitigate the effects of inter-core interference. Furthermore, coloring has inherent inefficiencies such as (i) precluding the use of superpages and (ii) increasing memory pressure which affects performance and predictability, as well as (iii) internal fragmentation (exclusively assigning 1 out of N colors, implicitly allocates 1/N$^{th}$ of physical memory, a portion of which may remain unused for small RTOSs or the SPH). While the latter could be solved by employing cache bleaching \cite{shanin2020} in heterogeneous platforms, to further minimize coloring bottlenecks, we advocate for SPH to adopt other proven, widely applicable contention mitigation mechanisms, e.g., bandwidth regulation mechanisms implemented via PMU-based CPU throttling \cite{Yun2013, Modica2018}. We also stress the importance of including support for hardware extensions such as Arm's Memory Partitioning and Monitoring (MPAM) \cite{arm2018mpam, Falk2021}, which provide flexible hardware means for partitioning cache space and memory bandwidth and call for platform designers to include such facilities in their upcoming designs targeting MCS. Finally, we stress the need for instrumentation, analysis, and profiling tools \cite{Sohal2020, Ghaemi2021} that integrate with these hypervisors to help system designers understand the trade-offs and fine-tune these mechanisms (e.g., through automation).

\mypara{Platform-Level Contention and Mitigation.} None of the studied SPH manages traffic from peripheral DMAs. We advocate that SPH must provide contention mitigation mechanisms at the platform level, e.g., (i) leveraging QoS hardware \cite{Sohal2020, Zini2022} available on the bus and (ii) controlling interference from DMA-capable devices or accelerators. Furthermore, since DMA masters still share SMMU structures (e.g., TLBs \cite{panchamukhi2015}), we hypothesize that bandwidth regulation techniques may fall short of efficiently mitigating interference at this level.

\mypara{Interrupt Injection Optimization.} Arm-based SPH's interrupt latency is mainly due to inadequate support in GICv2/3. GICv4 will provide direct interrupt injection support, but only for IPIs and MSIs. We want to raise awareness of Arm silicon makers and designers of the need for additional hardware support at the GIC level for direct injection of wired interrupts. The same holds for RISC-V \cite{Sa2021}.
Besides hardware support, we observed that simple SPH provide optimized interrupt injection paths. It is also possible to optimize this path in larger SPH (e.g., Xen) and in microkernels (e.g., moving injection logic to the microkernel). Finally, Bao and Jailhouse implement direct interrupt injection; however, we must stress that using this technique severely hinders the ability of the SPH to manage devices or implement any functionality dependent on interrupts. A plausible research direction would be a hybrid approach, i.e., selectively enabling direct injection only in specific cores for critical guests while providing the more complex functionality in cores running non-critical guests.

\mypara{Interrupt Priority Inversion Fix.} As discussed in Section \ref{sec:irq-latency}, the studied SPH suffer from partial interrupt priority inversion because all currently pending interrupts are handled by the hypervisor and injected in the guest before it can service the highest-priority one. We advocate for implementing a lightweight solution by dynamically setting the interrupt priority mask based on the priority of the last injected interrupt. This approach ensures the hypervisor only receives the next interrupt once the guest has handled the highest priority one.

\mypara{Critical VM Boot Priority.} Section \ref{sec:boot} highlights the issue of critical VM boot time overhead when booted under a dual-OS configuration. We advocate for the development of boot mechanisms that prioritize the boot of small critical VMs. However, as noted in Jumpstart \cite{Golchin2022}, it must encompass the full boot flow and be optimized across stages and components since the bottleneck of the boot time is in the image loading process performed by the bootloader, not the hypervisor.

\mypara{Per-Partition Hypervisor Replica.} Memory contention highly affects interrupt latency but can be minimized by assigning different colors for VMs and the hypervisor. Notwithstanding, coloring the hypervisor may prove wasteful and insufficient to address other interference channels internal to the hypervisor. We advocate for \textit{à la} multikernel \cite{baumann2009} implementations such as the one implemented in seL4, where the hypervisor image is replicated per cache partition \cite{Ge2019}, fully closing internal channels. For SPH with a small enough footprint, memory consumption or boot time costs should not be prohibitive.

\mypara{Architecture Flexibility.} Purely monolithic SPH (e.g., Jailhouse or Bao) have smaller code bases at the cost of feature richness and flexibility. The same holds for Xen, i.e., many widely-used rich features are absent when configured as an SPH (to minimize code size). On the other hand, the seL4 microkernel architecture is much more flexible as it allows for an isolated user space VMM per guest, providing more robust isolation and customization; however, it comes at the cost of non-negligible latencies. We advocate for novel architectures that combine microkernels' flexibility and strong fault encapsulation with SPH's simplicity and minimalist latencies by hosting per-partition VMMs directly at the hypervisor privilege level. Such a design could arguably be achieved by combining multikernel-like architectures \cite{baumann2009} and per-core memory protection mechanisms (e.g., Armv9 RME's GPT \cite{li2022}, or RISC-V PMP \cite{lee2020}) statically configured by firmware.

\mypara{Full IO Passthrough.} Pure static partitioning supports only passthrough IO. However, as highlighted by  \cite{Ramsauer2018}, there is a critical problem in providing full IO passthrough when controls over IO resources such as clock, reset, power, or pin-muxes cannot be securely partitioned or shared, e.g., if their MMIO registers reside on the same frame or they are configured via platform management co-processors oblivious of SPH's VMs. Thus, SPH should provide controlled guest access to these resources by emulation or through standard interfaces such as SCMI \cite{arm2022scmi}. Nevertheless, this would require including drivers in the hypervisor, increasing its code base. Again, we urge hardware designers to provide hardware primitives that enable SPH to pass through IO resource controls.


\vspace{-0.08cm}
\section{Related Work} \label{sec:related} 

There are several hypervisor analyses in the context of embedded and MCSs, but none provide a cross-section analysis and comparison on SPH. Some works focus on a single hypervisor while others evaluate a single metric or feature. In \cite{Patel2015}, authors compare the performance of Xvisor with Xen and KVM. Others have evaluated the effectiveness of cache coloring and bandwidth reservations in Xvisor \cite{Modica2018}. Similarly, in \cite{Kloda2019}, authors evaluate cache and DRAM bank coloring in Jailhouse. Other works have evaluated Jailhouse interrupt latency \cite{Pavic2018} or VM interference \cite{Danielsson2019}. There are also studies about the feasibility of using Xen and KVM as real-time hypervisors \cite{Abeni2020}, but mainly for x86. Little has been published regarding the new Xen Dom0-less and cache coloring features, but results can be found in \cite{Stabellini2020}. Evaluation of the seL4 CAmkES VMM has also been done for performance and interrupt latency \cite{millwood2020}. There have been works providing a qualitative analysis for MCS hypervisors, contrasting architectural approaches and highlighting future trends \cite{Cinque2022} while others layout guidelines on how to choose such a hypervisor in industrial settings \cite{Hamelin2020}.

\vspace{-0.08cm}
\section{Conclusion} \label{sec:concl} 

We have conducted the most comprehensive empirical evaluation of open-source SPH to date, focusing on key metrics for MCS. With that, we drew a set of observations that (i) will help industrial practitioners understand the trade-offs of SPH and (ii) raise awareness of the research and open-source communities to the still open problems in SPH. We are opening all artifacts to enable independent validation of results and encourage further exploration on SPH.

  


\section{Acknowledgments}

We would like to express our gratitude to the reviewers for their valuable feedback and suggestions, as well as to our friendly shepherd for guiding us in making final improvements. Additionally, we appreciate the time and thoughtful input from all the representatives of SPH, namely Ralf Ramsauer (Jailhouse), Stefano Stabellini (Xen), and Gernot Heiser (seL4/CAmkES-VMM). 
José Martins was supported by FCT grant SFRH/BD/138660/2018. This work is supported by FCT – Fundação para a Ciência e Tecnologia within the R\&D Units Project Scope UIDB/00319/2020, and European Union’s Horizon Europe research and innovation program under grant agreement No 101070537, project CROSSCON (Cross-platform Open Security Stack for Connected Devices).



\bibliographystyle{IEEEtran}
\bibliography{shedding-light.bib}




\end{document}